\documentclass[12pt,a4paper]{article}
\usepackage{jheplike,tocloft,ifpdf,tocloft,amsmath,mathrsfs,mathtools,xcolor}
\widowpenalty=500\clubpenalty=1000\unitlength=1mm\textheight=8.7in\hypersetup{pdftitle={},pdfcreator={},linkcolor=[rgb]{0.15,0.35,0.75},colorlinks=true,citecolor=[rgb]{0.675,0,0.2},urlcolor=[rgb]{0.15,0.35,0.65}}
\let\olditemize\itemize\renewcommand{\itemize}{\vspace{-2pt}\olditemize\setlength{\itemsep}{1pt}\setlength{\parskip}{0pt}\setlength{\parsep}{-0pt}}
\let\oldenumerate\enumerate\renewcommand{\enumerate}{\vspace{-4pt}\oldenumerate\setlength{\itemsep}{1pt}\setlength{\parskip}{0pt}\setlength{\parsep}{0pt}}
\makeatletter\renewcommand\section{\addtocontents{toc}{\protect\addvspace{-2.25\p@}}\@startsection {section}{1}{\z@}{0.5ex \@plus .2ex \@minus 0.2ex}{0.3ex \@plus.1ex\@minus .5ex}{\normalfont\large\bfseries}}
\renewcommand\subsection{\addtocontents{toc}{\protect\addvspace{-2.5\p@}}\@startsection {subsection}{1}{\z@}{0.5ex \@plus .2ex \@minus 0.2ex}{0.3ex \@plus.1ex\@minus .5ex}{\normalfont\bfseries}}
\renewcommand\subsubsection{\addtocontents{toc}{\protect\addvspace{-2.5\p@}}\@startsection {subsubsection}{1}{\z@}{0.5ex \@plus .2ex \@minus 0.2ex}{0.3ex \@plus.1ex\@minus .5ex}{\normalfont\bfseries}}

\newcommand{\eq}[1]{\vspace{-0.5pt}\begin{equation}#1\vspace{-0.5pt}\end{equation}}

\newcommand{\fwbox}[2]{\text{\makebox[#1][c]{$\hspace{-150pt}\displaystyle#2\hspace{-150pt}$}}}
\newcommand{\fwboxL}[2]{\text{\makebox[#1][l]{$#2$}}}
\newcommand{\fwboxR}[2]{\text{\makebox[#1][r]{$#2$}}}
\newcommand{\equivR}{\fwbox{14.5pt}{\hspace{-0pt}\fwboxR{0pt}{\raisebox{0.47pt}{\hspace{1.25pt}:\hspace{-4pt}}}=\fwboxL{0pt}{}}}
\newcommand{\equivL}{\fwbox{14.5pt}{\fwboxR{0pt}{}=\fwboxL{0pt}{\raisebox{0.47pt}{\hspace{-4pt}:\hspace{1.25pt}}}}}
\newcommand{\fig}[3]{\raisebox{#1}{\includegraphics[scale=#2]{#3}}}
\newcommand{\bigger}[1]{\raisebox{-0.95pt}{\scalebox{1.25}{$#1$}}}
\newcommand{\mi}{\raisebox{0.75pt}{\scalebox{0.75}{$\hspace{-0.5pt}\,-\,\hspace{-0.5pt}$}}}
\newcommand{\pl}{\raisebox{0.75pt}{\scalebox{0.75}{$\hspace{-0.5pt}\,+\,\hspace{-0.5pt}$}}}
\renewcommand{\phi}{\varphi}

\renewcommand{\hat}{\widehat}

\newcommand{\ab}[1]{\langle #1\rangle}

\newcommand{\newcap}{\mathrm{\raisebox{0.75pt}{{$\,\bigcap\,$}}}}
\newcommand{\tncap}{\scalebox{0.8}{$\!\newcap\!$}}
\newcommand{\x}[2]{{\color{black}(}\hspace{-0.85pt}{\color{black}#1}\hspace{-0.25pt}{\color{black}|}\hspace{-0.25pt}{\color{black}#2}\hspace{-0.85pt}{\color{black})}}
\newcommand{\Li}[2]{\hspace{1pt}\mathrm{Li}_{#1}(#2)}
\newcommand{\Zeta}[1]{\hspace{1pt}\zeta_{#1}}
\newcommand{\coupling}{{\color{highlight}a}}
\newcommand{\eps}{{\color{hblue}\delta}}
\newcommand{\higgsMass}{{\color{hblue}m_{{\color{hred}a}}^2}}
\newcommand{\higgsMassDull}{{\color{hblue}m_{{\color{black}a}}^2}}
\newcommand{\higgsRegName}{{\color{hblue}\text{Higgs}}}
\newcommand{\dciRegName}{{\color{hblue}\text{DCI}}}
\newcommand{\genericRegName}{{\color{hblue}\text{reg.}}}
\newcommand{\cyclicPrime}{\underset{\text{(delete duplicates)}}{\text{cyclic}_n}}
\newcommand{\cyclicPrimeN}[1]{\fwbox{38pt}{\underset{\text{(no dupl.)}}{\text{cyclic}_#1}}}
\newcommand{\mhvIntN}[5]{\Omega^{(#5)}\!\big[\hspace{-1pt}({\color{hblue}#1},\!{\color{hred}#2}),\!({\color{hblue}#3},\!{\color{hred}#4})\hspace{-1pt}\big]}
\newcommand{\mhvInt}[4]{\Omega\!\big[\hspace{-1pt}({\color{hblue}#1},\!{\color{hred}#2}),\!({\color{hblue}#3},\!{\color{hred}#4})\hspace{-1pt}\big]}
\newcommand{\mhvIntO}[4]{\Omega\!\big[\hspace{-1pt}({\color{hblue}#1},\!{\color{hblue}#2}),\!({\color{hred}#3},\!{\color{hred}#4})\hspace{-1pt}\big]}
\newcommand{\seedA}{{I}_1}\newcommand{\seedB}{{I}_2}\newcommand{\seedC}{{I}_3}\newcommand{\seedD}{{I}_4}\newcommand{\seedE}{{I}_5}
\newcommand{\seedAIntegrand}{\mathcal{I}_1}\newcommand{\seedBIntegrand}{\mathcal{I}_2}\newcommand{\seedCIntegrand}{\mathcal{I}_3}\newcommand{\seedDIntegrand}{\mathcal{I}_4}\newcommand{\seedEIntegrand}{\mathcal{I}_5}
\newcommand{\ancFileName}{\texttt{heptagon\_logarithm\_seed\_data.m}}
\definecolor{hblue}{rgb}{0,0,0.675}
\definecolor{hred}{rgb}{0.625,0.0,0.225}
\definecolor{dim}{rgb}{0.55,0.55,0.55}
\definecolor{highlight}{rgb}{0.0,0.445,0.4451}

\title{\texorpdfstring{~\\[0pt]{\LARGE \mbox{Conformally-Regulated Direct Integration}}\\{\LARGE of the Two-Loop Heptagon Remainder}\\[-24pt]}{Conformally Regulated Direct Integration of the Two-Loop Heptagon Remainder}}
\author[a,b,c]{\vspace{-20pt}Jacob~L.~Bourjaily,}\emailAdd{bourjaily@psu.edu}
\affiliation[a]{Niels Bohr International Academy and Discovery Center, Niels Bohr Institute,\\University of Copenhagen, Blegdamsvej 17, DK-2100, Copenhagen \O, Denmark}
\affiliation[b]{Center for the Fundamental Laws of Nature, Department of Physics,\\ Jefferson Physical Laboratory, Harvard University, Cambridge, MA 02138, USA}
\affiliation[c]{Institute for Gravitation and the Cosmos, Department of Physics,\\Pennsylvania State University, University Park, PA 16892, USA}
\author[a]{Matthias~Volk,}\emailAdd{mvolk@nbi.ku.dk}
\author[a]{Matt~von~Hippel}\emailAdd{mvonhippel@nbi.ku.dk}
\abstract{
We reproduce the two-loop seven-point remainder function in planar, maximally supersymmetric Yang-Mills theory by direct integration of conformally-regulated chiral integrands. The remainder function is obtained as part of the two-loop logarithm of the MHV amplitude, the regularized form of which we compute directly in this scheme. We compare the scheme-dependent anomalous dimensions and related quantities in the conformal regulator with those found for the Higgs regulator. 
}
\preprint{}

\begin{document}
\maketitle

\setcounter{page}{1}\vspace{10pt}%
\pagenumbering{arabic}
\vspace{6pt}\section{Introduction and Overview}\label{sec:introduction}\vspace{2pt}

The study of scattering amplitudes in recent decades has led to tremendous advances in both our understanding of quantum field theory and also our technical progress in computing the predictions made for experiment. Much of this progress can be attributed to the remarkable (and still surprising) simplicity of massless quantum field theories in four dimensions. Any such theory turns out to possess a connection to Grassmannian geometry \cite{ArkaniHamed:2009dn,ArkaniHamed:2009dg,ArkaniHamed:2012nw,Arkani-Hamed:2013jha} which has led to novel applications and greater understanding of perturbative amplitudes for an expanding class of quantum theories. This is true despite the subtlety involved in even defining the \mbox{$S$-matrix} for massless field theories! (But see~\cite{Frye:2018xjj,Hannesdottir:2019rqq} for recent progress on this problem.) 

Many of the difficulties of working with massless quantum field theories can be postponed by focusing on loop \emph{integrands} (`the sum of Feynman diagrams').  At the integrand level, there are several new and extremely powerful frameworks for expressing perturbative scattering amplitudes of an increasingly general class of theories. These tools include all-loop recursion relations \cite{ArkaniHamed:2010kv,Benincasa:2015zna}, bootstrap methods \mbox{\cite{Bourjaily:2011hi,Bourjaily:2015bpz,Bourjaily:2016evz}}, \mbox{$Q$-cuts} \cite{Baadsgaard:2015twa}, and the broad reach of generalized \mbox{\cite{Bern:1994zx,Bern:1994cg,Britto:2004nc,Bern:2006ew,Anastasiou:2006jv,Bern:2007ct,Cachazo:2008vp,Bern:2008ap,Berger:2008sj,Abreu:2017xsl}} and prescriptive \mbox{\cite{ArkaniHamed:2010gh,Bourjaily:2013mma,Bourjaily:2015jna,Bourjaily:2017wjl,Bourjaily:2018omh,Bourjaily:2019iqr,Bourjaily:2019gqu}} unitarity. 
It remains to be seen, however, how much of the simplicity of integrands can survive loop integration. 
Considering the extent to which the simplicity at the integrand-level arises specifically for theories of \emph{massless} particles in exactly \emph{four} dimensions, and that it is precisely these features that are responsible for infrared divergences whose regularization necessarily spoils them, it would not be surprising if much of this extra structure was lost to the infrared. Indeed, it would be reasonable to be skeptical that anything remarkable would be found for the actual infrared-safe quantities in which we are ultimately interested.

To test whether or not any of the niceness of amplitudes at the integrand-level survives the wrath and fury (the infrared regularization) of loop integration, it would be reasonable to simply `shut up and calculate'---by any means necessary---and see what emerges in the `[theoretical] data', so to speak. Of course, this will always be easier to accomplish for especially simple quantum field theories such as maximally supersymmetric ($\mathcal{N}\!=\!4$) Yang-Mills (`sYM') in the planar limit, for which the greatest computational leverage exists (largely due to this theory's special properties \cite{Mandelstam:1982cb,Brink:1982wv,Howe:1983sr,Drummond:2006rz,Alday:2007hr,Drummond:2008vq}).\\

There is a now-quite-famous example which illustrates what can be discovered through such a `compute first, understand later' strategy. It involves one of the simplest non-constant and non-trivial infrared-safe quantities in planar sYM: the (BDS) \emph{remainder function} for six particles at two-loop order. This quantity was determined through truly heroic efforts, first numerically \cite{Bern:2008ap} and then analytically \cite{DelDuca:2010zg}---in both cases, starting from an integrand-level expression obtained using unitarity-based methods; then regulating; then integrating. Within months of the publication of the analytic result, however, breathtaking simplicity was indeed found: the 18-page sum of hyperlogarithms in \cite{DelDuca:2010zg} could be written in a single line \cite{Goncharov:2010jf}! 

The ideas that led to the discovery of this simplicity would lead to a watershed of new and powerful techniques developed hand-in-hand with even greater evidence of simplicity surviving regularization and loop integration. Today, this particular quantity---the six-particle remainder function in planar sYM---is known to \emph{seven}(!) loops; and the seven-particle remainder is known (at least at `symbol-level') to four loops~\cite{Dixon:2011nj,Dixon:2013eka, Dixon:2014voa,Dixon:2014iba,Dixon:2015iva,Dixon:2016apl,Caron-Huot:2016owq,Caron-Huot:2019vjl,Caron-Huot:2019bsq,Drummond:2014ffa,Dixon:2016nkn,Drummond:2018caf}. Interestingly, after the two-loop result was found `the old fashioned way' in \cite{DelDuca:2010zg}---namely, by integrating Feynman integrands---all subsequent results were obtained using methods that \emph{made no reference to loop integrands or loop integration whatsoever}! While these ideas have more recently been applied to non-planar amplitudes in supersymmetric theories~\cite{Abreu:2018aqd,Chicherin:2018yne} and more broadly~\cite{Gehrmann:2015bfy,Badger:2017jhb,Abreu:2017hqn,Abreu:2018jgq,Badger:2018enw,Abreu:2018zmy,Chicherin:2018old,Chicherin:2018yne,Chicherin:2019xeg,Abreu:2019rpt}, they suffer from several fundamental limitations in applicability---in multiplicity, in the understanding (and simplicity) of the kinds of transcendental functions that arise in perturbation theory (including those described in e.g.~\cite{Bourjaily:2018yfy})---that prevent these ideas from rewriting the methods taught in textbooks, say.\\[-10pt]

One of the key motivations for our present work is the question of how much simplicity of loop integrands can be preserved through loop integration and regularization. Specifically, how can this bridge be crossed by \emph{direct} and general methods---without reference to any ansatz about the kinds of functions that may arise in particular cases. A key source of hope that a more direct (and therefore general) connection between the remarkable integrands for amplitudes in planar sYM \cite{Bourjaily:2013mma,Bourjaily:2015jna,Bourjaily:2017wjl} and the simple expressions that we now expect to find for infrared-safe quantities is the is the existence of the regulator introduced in \cite{Bourjaily:2013mma}, which allows infrared divergences to be regulated \emph{without breaking (dual-)conformal invariance}. Another critical source of optimism is the recent renaissance in in direct-integration technology for Feynman-parametric integrands~\cite{Brown:2009ta,Panzer:2014caa} (see also \cite{Bourjaily:2018aeq,Bourjaily:2019igt}). 

In this work, we test the robustness of this emerging bridge from integrands to integrals in the highly non-trivial case of the seven-point remainder function at two loops. This quantity was first determined at symbol-level in \cite{CaronHuot:2011ky} (see also \cite{CaronHuot:2010ek,CaronHuot:2011kk}), and later upgraded to a function-level result in \cite{Golden:2014xqf}. Here, we start from the chiral integrand representation for the logarithm of the amplitude given in \cite{ArkaniHamed:2010gh}, use the conformal regulator of \cite{Bourjaily:2013mma}, Feynman-parameterize these terms according to \cite{Bourjaily:2019jrk}, and integrate each piece using the technology of~\cite{Brown:2009ta,Panzer:2014caa}. The result is a novel (if not superior) representation of the two-loop remainder function, and a proof of concept that such a strategy can work. As a bonus, by combining this result with that of \cite{Bourjaily:2019jrk} for six particles, we are able to determine all of the scheme-dependent parts of the two-loop MHV-amplitude logarithm in the conformal regularization scheme.\\

This work is organized as follows. We start in section~\ref{sec:building_blocks} with a review of the the local integrands necessary for MHV amplitudes and their logarithms in planar sYM at two-loops and how these integrands can be regulated while preserving dual-conformal invariance. In section~\ref{sec:integration_details} we discuss how we can \emph{directly} integrate each of the integrands needed for the seven-particle logarithm, resulting in a representation in terms of explicit hyperlogarithmic functions. Our main results regarding the heptagon remainder function are described in section~\ref{sec:the_result}, where we determine the scheme-dependent parts of the logarithm of MHV amplitudes in the conformal regularization scheme and compare these with what is found for the Higgs regulator.\\

Available as part of this work's submission to the \texttt{arXiv}, we have prepared the ancillary file \ancFileName. This file contains: Feynman-parametric integrands for the five (cyclic) seeds which generate the seven-point logarithm at two loops; analytic expressions for each seed integral---given in terms of Goncharov hyperlogarithms---obtained via direct integration; details regarding the novel alphabets that arise for these integrals; and reference details regarding how our coordinates related to those used by \cite{Golden:2014xqf} in their representation of the two-loop heptagon remainder function.

\newpage\vspace{-6pt}\section{Local Integrands for (Logarithms of) MHV Amplitudes}\label{sec:building_blocks}\vspace{-0pt}

In this section, we give a rapid review of the representation (in terms of local Feynman integrals) of MHV amplitudes and their logarithms at two loops in the planar limit of sYM. In \mbox{\cite{ArkaniHamed:2010kv}} (see also the earlier work \cite{Bern:2008ap,Vergu:2009zm,Vergu:2009tu}), it was guessed (and checked) that the $n$-particle MHV amplitude integrand could be represented as\footnote{Notice that we have dropped the typical notation indicating N$^{(\mathbf{k}=0)}$MHV degree in `$\mathcal{A}_n^{(L)}$', as no other helicity sectors will be considered in this work.}
\vspace{-5pt}\eq{\mathcal{A}_n^{(L=2)}\equivR\frac{1}{2}\fwbox{40pt}{\bigger{\displaystyle\sum_{\substack{1\leq {\color{hblue}a}\leq n\\{\color{hblue}a}<{\color{hblue}b}<{\color{hred}c}<\\{\color{hred}d}<n+{\color{hblue}a}}}}}\hspace{-10pt}\fig{-29.67pt}{1}{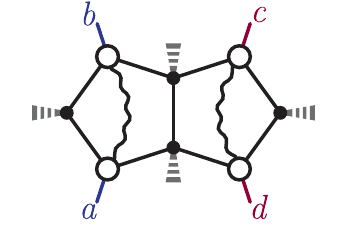}\hspace{-2pt},\label{mhv_two_loop_integrand}}
where the double-pentagons, herein `$\mhvIntO{a}{b}{c}{d}$', have precise loop-dependent numerators (indicated by the wavy-lines in the figure) expressed in terms of momentum twistors~\cite{Hodges:2009hk}:
\begin{align}\hspace{-7.5pt}\fig{-29.67pt}{1}{mhv_general_double_pentagon}\hspace{-7.5pt}&\equivL\mhvIntO{a}{b}{c}{d}\label{mhv_double_pentagon_formula}\\[-8pt]
&\equivR\frac{\ab{(\hspace{-1pt}\ell_1\hspace{-1pt})({\color{hblue}a\mi\!1\,a\,a\pl\!1})\tncap({\color{hblue}b\mi\!1\,b\,b\pl\!1})}\ab{{\color{hblue}ba}\,{\color{hred}dc}}\ab{(\hspace{-1pt}\ell_2\hspace{-1pt})({\color{hred}c\mi\!1\,c\,c\pl\!1})\tncap({\color{hred}d\mi\!1\,d\,d\pl\!1})}}{\x{\ell_1}{{\color{hblue}a}}\x{\ell_1}{{\color{hblue}a\pl\!1}}\x{\ell_1}{{\color{hblue}b}}\x{\ell_1}{{\color{hblue}b\pl\!1}}\x{\ell_1}{\ell_2}\x{\ell_2}{{\color{hred}c}}\x{\ell_2}{{\color{hred}c\pl\!1}}\x{\ell_2}{{\color{hred}d}}\x{\ell_2}{{\color{hred}d\pl\!1}}}.\nonumber
\end{align}
As usual, we are using the notations $\x{a}{b} \equivR (x_a - x_b)^2$ where $x_a$ are the dual coordinates related to the momenta through $p_a\equivL x_{a+1}\mi x_{a}$, and $\ab{a b c d} \equivR \det(z_a ,\, z_b ,\, z_c ,\, z_d)$ for the ordinary four-brackets of momentum twistors.

We should clarify that the factor of `$1/2$' appearing in (\ref{mhv_two_loop_integrand}) is really a \emph{symmetry factor}: it accounts for the fact that the summand includes each contribution exactly twice---provided we view the integrand in (\ref{mhv_double_pentagon_formula}) as being (implicitly) symmetrized with respect to $\ell_1\leftrightarrow\ell_2$; in particular, this factor of $1/2$ could be dispensed by an instruction to `delete duplicates' from the RHS (something often left implicit in the relevant literature). As $\mhvIntO{a}{b}{c}{d}$ and $\mhvIntO{{\color{hred}c}}{{\color{hred}d}}{{\color{hblue}a}}{{\color{hblue}b}}$ are identical upon integration, \emph{we consider them equivalent} (a.k.a.\ `duplicates')---a potential source of confusion below, for which we apologize.

Notice that the definition of $\mhvIntO{a}{b}{c}{d}$ depends on up to twelve momentum twistors
\eq{\{z_{{\color{hblue}a-1}},z_{{\color{hblue}a}},z_{{\color{hblue}a+1}}\}\bigger{\cup}\{z_{{\color{hblue}b-1}},z_{{\color{hblue}b}},z_{{\color{hblue}b+1}}\}\bigger{\cup}\{z_{{\color{hred}c-1}},z_{{\color{hred}c}},z_{{\color{hred}c+1}}\}\bigger{\cup}\{z_{{\color{hred}d-1}},z_{{\color{hred}d}},z_{{\color{hred}d+1}}\}\,,}
with cyclic labeling understood. Especially for low multiplicity, these indices can overlap considerably. When it is necessary to disambiguate the multiplicity $n$, implicit in the definition (\ref{mhv_double_pentagon_formula}) above, we will signify this by writing `$\mhvIntN{{\color{black}a}}{{\color{black}b}}{{\color{black}c}}{{\color{black}d}}{{\color{hred}n}}$'. 

Shortly after the formula (\ref{mhv_two_loop_integrand}) appeared in \mbox{\cite{ArkaniHamed:2010kv}}, a similar expression was derived in \mbox{\cite{ArkaniHamed:2010gh}} for the four-dimensional integrand of the two-loop \emph{logarithm} of the MHV amplitude,
\eq{\hspace{-80pt}\log\!\big(\mathcal{A}_n\big)^{(L=2)}=\mathcal{A}_n^{(L=2)}-\frac{1}{2}\Big(\mathcal{A}_n^{(L=1)}\Big)^2=-\frac{1}{4}\hspace{-5pt}\fwbox{40pt}{{\displaystyle\sum_{\substack{1\leq {\color{hblue}a}<n\\{\color{hblue}a}<{\color{hblue}c}<{\color{hred}b}<\\{\color{hred}d}<n+{\color{hblue}a}}}\hspace{0pt}\fwboxL{0pt}{\hspace{-3.5pt}\mhvInt{a}{b}{c}{d}\,.}}}\hspace{0pt}
\label{mhv_two_loop_logarithm_integrand}}
(As before, the factor of `$1/4$' above is merely a symmetry factor: the appropriate prefactor would be 1 times each term in the summand \emph{without duplication}.) Notice that the summand in (\ref{mhv_two_loop_logarithm_integrand}) now excludes the possibility that ${\color{hblue}a}\!+\!1\!={\color{hred}b}$ and---more importantly---the summand requires that ${\color{hblue}c}\!\in\!\{{\color{hblue}a} \pl1,\!\ldots,\!{\color{hred}b} \mi1\}$. 

It is instructive to see a few instances of equation (\ref{mhv_two_loop_logarithm_integrand}). Without symmetry factors, but being explicit about the fact that cyclic seeds should be summed only \emph{without duplication}, and being very careful about which cyclic seeds necessitate clarification about when multiplicity matters, the two-loop logarithms of MHV amplitudes for 4-8 particles are as follows:
\eq{\hspace{-84pt}\log\!\big(\hspace{-1pt}\mathcal{A}_{4}\hspace{-1pt}\big)^{(2)}\!\!\!=\!-\!\Bigg[\fwboxL{300pt}{\mhvIntN{2}{4}{3}{1}{4}\!+\!\cyclicPrimeN{4}\Bigg]=-\mhvIntN{2}{4}{3}{1}{4}\,,}\hspace{-50pt}\label{4pt_log_integrand}}
\eq{\hspace{-84pt}\log\!\big(\hspace{-1pt}\mathcal{A}_{5}\hspace{-1pt}\big)^{(2)}\!\!\!=\!-\!\Bigg[\fwboxL{300pt}{\mhvIntN{2}{4}{3}{5}{5}\!+\!\cyclicPrimeN{5}\Bigg]=-\Bigg[\hspace{1.2pt}\mhvIntN{2}{4}{3}{5}{5}\!+\!\text{cyclic}_5\,\hspace{1.2pt}\Bigg]\,,}\hspace{-50pt}\label{5pt_log_integrand}}
\eq{\hspace{-84pt}\log\!\big(\hspace{-1pt}\mathcal{A}_{6}\hspace{-1pt}\big)^{(2)}\!\!\!=\!-\!\Bigg[\fwboxL{300pt}{\mhvInt{2}{4}{3}{5}\!\!+\!\mhvIntN{2}{4}{3}{6}{6}\!\!+\!\mhvIntN{2}{5}{3}{6}{6}\!+\!\cyclicPrimeN{6}\Bigg]\,,}\hspace{-50pt}\label{6pt_log_integrand}}
\eq{\begin{split}
\hspace{-90pt}\log\!\big(\hspace{-1pt}\mathcal{A}_{7}\hspace{-1pt}\big)^{(2)}\!\!\!=\!-\!\Bigg[&\fwboxL{300pt}{\mhvInt{2}{4}{3}{5}\!\!+\!\mhvInt{2}{4}{3}{6}\!\!+\!\mhvInt{2}{5}{3}{6}}\hspace{-41.5pt}\\
&\hspace{-12.125pt}+\!\mhvIntN{2}{4}{3}{7}{7}\!\!+\!\mhvIntN{2}{5}{3}{7}{7}\!+\!\cyclicPrimeN{7}\Bigg]\,,\end{split}\label{7pt_log_integrand}}
\eq{\begin{split}
\hspace{-89pt}\log\!\big(\hspace{-1pt}\mathcal{A}_{8}\hspace{-1pt}\big)^{(2)}\!\!\!=\!-\!\Bigg[&\fwboxL{300pt}{\mhvInt{2}{4}{3}{5}\!\!+\!\mhvInt{2}{4}{3}{6}\!\!+\!\mhvInt{2}{5}{3}{6}}\hspace{-45pt}\\
&\hspace{-12.125pt}+\!\mhvInt{2}{4}{3}{7}\!\!+\!\mhvInt{2}{5}{3}{7}\!\!+\!\mhvInt{2}{6}{3}{7}\!\!+\!\mhvIntN{2}{4}{3}{8}{8}\hspace{-50pt}\\
&\hspace{-12.125pt}+\!\mhvIntN{2}{5}{3}{8}{8}\!\!+\!\mhvIntN{2}{5}{4}{8}{8}\!\!+\!\mhvIntN{2}{6}{4}{8}{8}\!+\!\cyclicPrimeN{8}\Bigg]\,.\hspace{-53.35pt}\end{split}\label{8pt_log_integrand}}

There are a couple of things to notice about these representations. First, observe that for more than six particles the majority of cyclic seeds can be chosen to be independent of $n$; therefore, these contributions remain unchanged beyond some threshold multiplicity. The second thing to notice is that it is fairly easy to organize contributions according to their degrees of infrared divergence:\footnote{In dimensional regularization, `$\log^k$-divergent' should be understood as `$1/\epsilon^k$-divergent'.}
\eq{\begin{split}
    \log^2\text{-}&\text{divergent: }\mhvInt{2}{4}{3}{5}\,\text{ \emph{only},}\\
    \log^1\text{-}&\text{divergent: }\mhvInt{2}{4}{3}{b}\,\text{ for }{\color{hred}b}>5,
    \end{split}
  }
with all other integrals finite. In particular, notice that the \emph{only} cyclic seed with a $\log^2$-divergence is $\mhvInt{2}{4}{3}{5}$ and that this integral is $n$-independent once it is evaluated for any $n\geq6$. We will return to the consequences of this fact momentarily.

To regulate these divergences, we employ the so-called `dual-conformal' regularization scheme introduced in \mbox{\cite{Bourjaily:2013mma}}, wherein each (massless) external particle is taken off the lightcone by an amount proportional to the conformally-invariant parameter denoted `$\eps$' according to
\eq{p_a^2\mapsto p_a^2+\eps\frac{(p_{a-1}+p_a)^2(p_a+p_{a+1})^2}{(p_{a-1}+p_a+p_{a+1})^2} = \x{a}{a+1}+\eps\frac{\x{a-1}{a+1}\x{a}{a+2}}{\x{a-1}{a+2}}\,.}
(There is an alternative definition of this regulator expressed in terms of dual-momentum coordinates---where each dual coordinate $x_a$ is shifted by a small amount in the direction of its cyclic neighbor, $x_{a+1}$; these two definitions are not identical for finite $\eps$, but they result in regulated integrals equivalent to $\mathcal{O}(\eps)$.)

\vspace{-0pt}\subsection{Specific Contributions  to the Seven-Point Logarithm}\label{subsec:seven_point_details}\vspace{-0pt}
%

As seven particles is the primary example of interest to us here, it is worthwhile to give the five cyclic generators in (\ref{7pt_log_integrand}) individual names. Let us therefore define
\eq{\begin{array}{@{}c@{}}\seedAIntegrand\equivR\mhvInt{2}{4}{3}{5}\,,\quad \seedBIntegrand\equivR\mhvInt{2}{4}{3}{6}\,,\quad \seedCIntegrand\equivR\mhvInt{2}{5}{3}{6}\,,\\[4pt]
\seedDIntegrand\equivR\mhvIntN{2}{4}{3}{7}{7}\,,\qquad\seedEIntegrand\equivR\mhvIntN{2}{5}{3}{7}{7}\,.\end{array}\label{heptagon_log_seeds}}
Notice that from our discussion above, only $\seedAIntegrand$ will be $\log^2$-divergent in the infrared upon integration, while $\{\seedBIntegrand,\seedDIntegrand\}$ will be $\log^1$-divergent; the two seeds $\{\seedCIntegrand,\seedEIntegrand\}$ are infrared finite, and therefore do not require any regularization.

We will discuss how each of the contributions (\ref{heptagon_log_seeds}) can be evaluated in the following section. But already now we can observe an important consequence of the fact that $\seedAIntegrand$ depends exclusively on momentum twistors $\{z_1,\ldots,z_6\}$: its evaluation will be the same for seven particles as it was for six. More specifically, $\seedAIntegrand$ is essentially identical to what was computed (as part of what was called `$\mathcal{I}_{15}$') in \mbox{\cite{Bourjaily:2019jrk}}
\begin{align}I_1&\equivR\!\! \int\!\!\!d^4\!\ell_1d^4\!\ell_2\;\mathcal{I}_1\label{seedAintegral_formula}\\
&=\frac{1}{4}\Bigg[2\Zeta{2}\log^2\!(\eps)\pl 6\Zeta{3}\Big[\!\log\!(\eps)\pl 1\!\Big]-\Zeta{2}^2\mi 2\Zeta{2}G_{0,1}(1\mi w)\pl G_{0,0,0,1}(1\mi w)\mi G_{0,1,0,1}(1\mi w)\Bigg]\,,\nonumber\end{align}
where
\eq{w\equivR\frac{\x{{\color{hred}3}}{{\color{hblue}5}}\x{{\color{hblue}6}}{{\color{hred}2}}}{\x{{\color{hred}3}}{{\color{hblue}6}}\x{{\color{hblue}5}}{{\color{hred}2}}}=\frac{\ab{{\color{hred}23}\,{\color{hblue}45}}\ab{{\color{hblue}56}\,{\color{hred}12}}}{\ab{{\color{hred}23}\,{\color{hblue}56}}\ab{{\color{hblue}45}\,{\color{hred}12}}}.}
Notice that we are reserving calligraphic symbols to denote \emph{integrands} and \emph{italic} symbols to indicate \emph{integrals}.

As $\seedAIntegrand$ is the only cyclic seed with a $\log^2$-divergence for arbitrary $n$, it is wholly responsible for the leading divergence of the logarithm of MHV amplitudes at two loops. The coefficient of this divergence is related to the (scheme independent) \emph{cusp anomalous dimension}, and the attentive reader can already see that (\ref{seedAintegral_formula}) captures the right behavior. We will see this in detail in \mbox{section \ref{sec:the_result}} below; but before we do, it is worthwhile to describe how the other seven-point seeds have been evaluated analytically.

\vspace{-0pt}\section{Feynman Parameterization and Direct Integration}\label{sec:integration_details}\vspace{-0pt}

Following the strategy described in \mbox{\cite{Bourjaily:2019jrk}}, it is straightforward to Feynman-parameterize and regulate each of the contributions (\ref{heptagon_log_seeds}). For each of the double-pentagon integrals, this will result in a rational, five-dimensional parametric integral representation of the form\footnote{We hope the reader will forgive our abuse of notation in using `$\mathcal{I}_i$' to denote both the loop-momentum-space and Feynman-parametric integrands.}
\eq{I_i\equivR\int\limits_0^\infty\!\!\!\big[d^3\!\vec{\alpha}\big]\!d^2\!\vec{\beta}\;\;\mathcal{I}_i\big(\vec{\alpha},\vec{\beta};\{z_1,\ldots,z_7\},\eps\big)\label{schematic_parametric_rep}}
In the integral above, $\big[d^3\!\vec{\alpha}\big]\equivR d^4\vec{\alpha}\,\,\delta\big(\alpha_j\mi1\big)$ (for any $j$) represents a projective, 3-dimensional volume-form; while the $\beta$ integrations are not taken to be projective. This distinction is largely irrelevant due to the Cheng-Wu theorem~\cite{Cheng:1987ga}; but it reflects the way in which the parametric representations were derived via \cite{Bourjaily:2019jrk}, and we find it useful to keep this information. In the ancillary file, we provide a parametric representation of each of the seven-point integrals in (\ref{heptagon_log_seeds}).

\newpage\vspace{0pt}\subsection{(Cluster) Coordinate Charts for Heptagon Integrals}\label{subsec:cluster_coordinates}\vspace{-0pt}

In~\eqref{mhv_double_pentagon_formula} we have given the formula for $\mhvInt{a}{b}{c}{d}$ in terms of momentum twistors $z_a\!\in\!\mathbb{P}^3$ for $a = 1, \ldots, n$ that parameterize the kinematic space of $n$ massless particles.
As described in detail in \cite{Bourjaily:2018aeq} a momentum-twistor parameterization is preferred over one expressed in terms of dual-momentum $x$-coordinates, as twistor space immediately provides us with an integrand that is rational in terms of an independent set of conformal variables.

It turns out that the default cluster coordinates on $G_+(4,n)$ of the {\sc Mathematica} package \texttt{positroids} \cite{Bourjaily:2012gy} provide a very convenient chart for our present purposes. For a more detailed discussion of these coordinates we again refer the reader to \mbox{\cite{Bourjaily:2018aeq}}. For seven points, we can think of these coordinates as parameterizing seven momentum twistors $Z\equivL (z_1 \cdots z_7)$ according to 
\eq{Z(\{e_a^i\})\equivR\raisebox{-0pt}{$\left(\raisebox{28pt}{}\right.$}\begin{array}{@{}c@{$\;\;$}c@{$\;\;$}c@{$\;\;$}c@{$\;\;$}c@{$\;\;$}c@{$\;\;$}c@{}}1&1\pl e_6^3\pl e_7^3&e_6^3\pl (1\pl e_6^2)e_7^3&e_6^2e_7^3&0&0&0\\[-2pt]0&1&1\pl e_6^2\pl e_7^2&e_6^2\pl(1\pl e_6^1)e_7^2&e_6^1e_7^2&0&0\\[-2pt]0&0&1&1\pl e_6^1\pl e_7^1&e_6^1\pl e_7^1&e_7^1&0\\[-2pt]0&0&0&1&1&1&1\end{array}\raisebox{-0pt}{$\left.\raisebox{28pt}{}\right)$}\,;\label{seven_point_parameterized_twistors}}
or, if viewed as coordinates (maps from $G_+(4,7)\mapsto\mathbb{R}^6$), the parameters $\{e_a^i\}$ correspond to the conformal cross-ratios
\vspace{-18pt}\eq{\hspace{-15pt}\fwbox{0pt}{\begin{array}{@{}l@{$\;$}l@{$\;$}l@{}}~\\[14pt]\displaystyle e_6^1\equivR\frac{\ab{1234}\ab{1256}}{\ab{1236}\ab{1245}},\,&\displaystyle e_6^2\equivR\frac{\ab{1235}\ab{1456}}{\ab{1256}\ab{1345}},\,&\displaystyle e_6^3\equivR\frac{\ab{1245}\ab{3456}}{\ab{1456}\ab{2345}},\\[8pt]
\displaystyle e_7^1\equivR\frac{\ab{1234}\ab{1235}\ab{1267}}{\ab{1236}\ab{1237}\ab{1245}},\,&\displaystyle e_7^2\equivR\frac{\ab{1236}\ab{1245}\ab{1567}}{\ab{1256}\ab{1267}\ab{1345}},\,&\displaystyle e_7^3\equivR\frac{\ab{1256}\ab{1345}\ab{4567}}{\ab{1456}\ab{1567}\ab{2345}}.\end{array}}\label{seven_point_seed_coordinate_chart}}
%

\vspace{-0pt}\subsection{Divide and Conquer: Parametric Integration via Various Pathways}\label{subsec:integration_tricks}\vspace{-0pt}

The seed integrands expressed in this way can be integrated in terms of hyperlogarithms~\cite{Anastasiou:2013srw,Panzer:2014gra,Brown:2008um} (e.g.\ using \texttt{HyperInt}~\cite{Panzer:2014caa}) if there exists an order of the integration variables in which the integrand is linearly reducible. Na\"ively, however, this turns out not to be the case for any of the integrals at hand: all require some minor `tricks' of integration analogous to those discussed in, for example, \mbox{\cite{Panzer:2014gra,Bourjaily:2018aeq,Besier:2018jen,Bourjaily:2019jrk,Bourjaily:2019igt}}. 

Among the integration techniques required are those that allow us to extract the leading terms in the limit of $\eps\!\to\!0^+$ (for the integrals which require regularization). We were able to effectively use the methods discussed in \mbox{\cite{Bourjaily:2019jrk}}; we refer the reader to appendix B.1 and the ancillary files of that work for a more thorough explanation and illustrative examples. 

Of the two infrared finite integral seeds, only $\seedEIntegrand$ required mild cleverness to integrate directly. For this integral, a strategy which started along similar lines to that described in \mbox{\cite{Bourjaily:2019igt}} worked quite well. Specifically, starting from the Feynman-parametric integrand representation of the form (\ref{schematic_parametric_rep}) (provided in the ancillary file), we found that the integrals over $\alpha_2,\beta_1,$ and $\beta_2$ could each be performed rationally---i.e.\ without introducing any algebraic dependence on the remaining integration variables in the arguments of the hyperlogarithms or their prefactors. 

The (projective) two-fold parametric representation of $\seedEIntegrand$ obtained in this way suffers from a mild problem all-too familiar in these examples: integration in any one of the remaining variables would result in some terms with a square root depending (quadratically) on the final integration variable. Such an obstruction is easy to overcome by changing variables (Euler substitution) as described in e.g.\ \mbox{\cite{Panzer:2014gra,Besier:2018jen}}. But a better pathway to integration turns out to exist: the individual terms of the two-fold parametric representation of $\seedEIntegrand$ can be divided into groups which separately avoid this issue with respect to integration in $\alpha_4$ or $\alpha_1$. This results in a final expression with fewer `spurious' algebraic symbol letters---to be discussed in the next section.

\vspace{-0pt}\subsection{Refining the Results of Integration (Removing Spurious Letters)}\label{subsec:processing_integrals}\vspace{-0pt}

Following the strategies discussed above, it was fairly easy to obtain hyperlogarithmic (regulated, if necessary) expressions for integrals $\{\seedA,\ldots,\seedD\}$; but integration of $\seedEIntegrand$ required some cleverness, resulting in a representation of $\seedE$ that is considerably more complicated in two key aspects: first, the representation we obtained for $\seedE$ was not manifestly \emph{pure} in the sense of~\cite{ArkaniHamed:2010gh,Arkani-Hamed:2014via}---namely, it was expressed as a sum of hyperlogarithms with non-constant (algebraic) coefficients; and second, it was expressed in terms of hyperlogarithms with many (suspected to be `spurious') algebraic branch points. Let us discuss each of these complications in turn. 

The first complication, regarding the non-manifest `purity' of $\seedE$ turns out to be straightforward to deal with. First, we should clarify why we expected $\seedE$ to be pure despite its representation. Although the conformal regulator is known to spoil an integrand's purity (see the discussion in \mbox{\cite{Bourjaily:2019jrk}}), we strongly expect the logarithm of the amplitude (the cyclic sum of all seeds) to be pure; as  $\{\seedA,\ldots,\seedD\}$ were individually pure, it would require considerable magic for impurities of $\seedE$ to cancel amongst themselves in the cyclic sum.

Setting aside our expectations about $\seedE$'s purity, it turns out to be fairly easy to test whether or not any non-manifestly pure sum of hyperlogarithms is in fact pure. Suppose that some non-manifestly pure sum of hyperlogarithms $I(\{e_a^i\})$ depending on parameters $\{e_a^i\}$ is in fact pure; then we should be able to re-express it in terms of some \emph{basis} of hyperlogarithms $\{G_\beta\}$:
\eq{I(\{e_a^i\})\equivR\sum_\alpha R_\alpha\!(\{e_a^i\}) G_\alpha\!(\{e^i_a\}) \Rightarrow \sum_\beta c_\beta G_\beta\!\left(\{e^i_a\}\right)\,,\label{purported_identity}}
where $R_\alpha$ are rational(/algebraic)-function prefactors, $c_\beta$ are constants, and $G_\alpha$, $G_\beta$ multiple polylogarithms. In order for (\ref{purported_identity}) to be true, there would need to be some relations among the functions $G_\alpha$. Crucially, any such relations would necessarily be linear and have \emph{constant} coefficients---as all relations between multiple polylogarithms are expected to preserve transcendental weight and not involve any rational functions of their arguments. 

Now suppose we were to Taylor-expand each coefficient $R_\alpha$ in (\ref{purported_identity}) around some point $\hat{e_a^i}$ where all the $R_\alpha$'s are non-singular. Then we would have
\eq{\sum_\alpha \left[\sum_{j=0}^{\infty}R_\alpha^{(j)}\Big(e_a^i-\hat{e_a^i}\big)^{j}\right]G_\alpha\!(\{e^i_a\}) = \sum_\beta c_\beta G_\beta\!\left(\{e^i_a\}\right)\,.\label{series_version_of_identity}}
Since all purported relations among the $\{G_\alpha\}$ are linear, this requires that the identity (\ref{series_version_of_identity}) holds for each term in the Taylor series separately. In particular, it must hold at leading order. Moreover, as each $R_{\alpha}^{(0)}$ is just some constant, this term in the left-hand side of (\ref{series_version_of_identity}) is itself pure. 

The above discussion shows that \emph{when an integral is in fact pure}, any representation like that on the LHS of (\ref{purported_identity}) can be replaced by series-expanding each coefficient to leading order around any non-singular point, resulting in a manifestly pure representation. To test whether or not an integral is in fact pure, we can simply evaluate both ends of this algorithm numerically and check that they agree. For $\seedE$ we have checked in this way that it is in fact pure, and have provided a manifestly pure representation (obtained in this way) in the ancillary file.\\[-10pt]

The second complication about the representation of $\seedE$ obtained in the manner described above (namely, divide and conquer) is that this method has a tendency to introduce `spurious' branch points among terms (which cancel between the divided pieces). When these spurious branch points are not rational in the variables $\{e_a^i\}$, we know of no general strategy to canonically eliminate them (as we would by choosing a fibration basis, for example, had they been rational). Removing a dependence on spurious square roots from polylogarithmic expressions is in general a difficult problem, and one we will not attempt to solve here. 

Although we have not found a representation for $\seedE$ free of spurious square-root branch points, we are able to confirm that all non-rational branch points are indeed spurious. To do this, we first compute the symbol~\cite{Goncharov:2010jf,Duhr:2011zq} of $\seedE$, resulting in an alphabet of 85 letters, 22 of which involve square roots. These algebraic letters appear in pairs of the form $\rho\pm\sqrt{\sigma}$, which can be multiplied to generate root-free letters, leaving us with only 11 algebraic letters to analyze. 

These 11 spurious letters are not all independent. Unlike for symbols involving only rational letters, merely factoring square-root letters is not enough to trivialize all identities due to the absence of a unique factorization domain (for further discussion, see \cite{Bourjaily:2019igt}). Here we do not need to make use of the more mathematically sophisticated methods \cite{Bourjaily:2019igt}. Instead, we simply observe that products of pairs of our remaining eleven letters can yield letters that appear elsewhere in the symbol. By taking into account all such pairings, we find six relations between the 11 letters, and imposing these results in a manifestly rational symbol. This rationalized symbol for $\seedE$ can now be viewed as canonical, and consists of 47 letters (functions of momentum twistor cross-ratios).\\[-10pt]

From the symbol of $\seedE$, it would be possible to \emph{reconstruct} a rational, hyperlogarithmic representation---using essentially the same techniques by which the two-loop heptagon remainder function was first obtained in \cite{Golden:2014xqf} from its symbol, which in turn was first computed in \cite{CaronHuot:2011ky} (see also \cite{CaronHuot:2010ek,CaronHuot:2011kk}). We choose not to pursue this for $\seedE$ because functional \emph{reconstruction} is not our goal here. Rather, we are interested in how far we may push \emph{direct} integration of local integrals. One can easily check that the representation we give for $\seedE$---despite its spurious letters---perfectly matches Monte Carlo integration.

\vspace{-0pt}\section{The Two-Loop Heptagon Remainder Function}\label{sec:the_result}\vspace{-0pt}

We are now ready to describe the results of our analysis---to discover the form of the (all-orders) relationship between the logarithm of the MHV amplitude and the so-called `BDS' remainder function \cite{Bern:2008ap} in the conformal regularization scheme. Both for the sake of comparison and in order to introduce some useful notation, let us first pause to review the form of this relationship in the so-called `Higgs' regularization scheme described in \cite{Alday:2009zm,Henn:2010ir}.

\vspace{-0pt}\subsection{{\it Exempli Gratia}: Higgs-Regulated (Logarithms of) MHV Amplitudes}\label{subsubsec:higgs_regulator_reference}\vspace{-0pt}

At leading order in the coupling $\coupling\equivR g^2N_c/(8\pi^2)$, the MHV amplitude (divided by the tree) and its logarithm are identical (in any regularization scheme `$\genericRegName$'):
\eq{\log\!\big(A_{n,\genericRegName}\big)\equivL\sum_{\ell=1}^{\infty}\coupling^\ell\log\!\big(A_{n,\genericRegName}\big)^{(\ell)}=\coupling A_{n,\genericRegName}^{(1)}+\coupling^2\Big[A_{n,\genericRegName}^{(2)}-\frac{1}{2}\big(A_{n,\genericRegName}^{(1)}\big)^2\Big]+\mathcal{O}(\coupling^3)\,.\label{generic_log_expansion}}
(Recall our convention that calligraphic symbols such as $\mathcal{A}$ denote \emph{integrands} while italic symbols such as $A$ denote \emph{integrals}.)
As such, it is useful to first review the form of the one-loop amplitude in the relevant regularization scheme. 

For the Higgs regulator described in \cite{Alday:2009zm,Henn:2010ir}, one loop MHV amplitudes take the form 
\eq{A_{n,\higgsRegName}^{(1)}\equivL -\frac{1}{4} \Bigg[\sum_{{\color{hred}a}=1}^n\log^2\!\left(\frac{\higgsMass}{\x{{\color{hred}a}}{{\color{hred}a+2}}}\right)\Bigg]+F_{n,\higgsRegName}^{(1)}+\mathcal{O}(\higgsMass)\,,\label{original_mass_regulator}}
where $F_{n,\higgsRegName}^{(1)}$ is the so-called\footnote{It is so-called despite the fact that the leading term of (\ref{original_mass_regulator}) includes parts finite as $\higgsMassDull\!\to\!0$.} `finite part' of the one-loop amplitude in this scheme, and where we have added an index `${\color{hred}a}$'$\in[n]$ to distinguish between the various internal masses $\higgsMass$ (which are typically taken to be the same). Notice that we are using dual-momentum notation where $\x{a}{b}\equivR (x_{a} - x_{b})^2\equivR(p_a+\ldots+p_{b-1})^2$. It is worthwhile to consider the direction along the Higgs branch where these masses scale according to
\eq{\higgsMassDull\mapsto\eps\frac{\x{{\color{black}a-1}}{a+1}\x{a}{a+2}}{\x{a-1}{a+2}}\,\label{leg_dependent_mass_rule}}
under which 
\eq{A_{n,\higgsRegName}^{(1)}\underset{\text{(\ref{leg_dependent_mass_rule})}}{\longmapsto} -\frac{1}{4} \Bigg[n\log^2\!(\eps)+\log(\eps)\log(w_1\cdots w_n)+\sum_{a=1}^n\log^2\!\left(\!\frac{\x{a}{a+2}}{\x{a}{a+3}}\!\right)\Bigg]+F_{n,\higgsRegName}^{(1)}+\mathcal{O}(\eps) \,,\label{conformal_like_version_of_higgs_one_loop}}
where the cross-ratio $w_a$ is given by
\eq{w_a\equivR\frac{\x{a}{a+2}\x{a+3}{a+5}}{\x{a}{a+3}\x{a+2}{a+5}}\,.\label{w_cross_ratios_defined}}
This is extremely similar to the form of the one loop amplitude in the conformal regularization scheme. Before we get to that, however, let us first recall a few more facts about the Higgs regulator and the form that the logarithm (\ref{generic_log_expansion}) takes in this scheme. 

In \mbox{\cite{Henn:2010ir}}, the all-order form of the logarithm (\ref{generic_log_expansion}) was represented according to the BDS ansatz \cite{Bern:2005iz} as 
\eq{\begin{split}\hspace{-10pt}\log(A_{n,\higgsRegName})\equivL&-\frac{\gamma_c(\coupling)}{16}A_{n,\higgsRegName}^{(1)}\!+\!\frac{\widetilde{\cal G}_0(\coupling)}{2}\sum_{{\color{hred}a}=1}^n\log\!\left(\!\frac{\higgsMass}{\x{{\color{hred}a}}{{\color{hred}a+2}}}\!\right)+n\widetilde{f}(\coupling)+\widetilde{C}(\coupling)+R_n(\coupling)\hspace{-20pt}\\
&\fwboxR{0pt}{\underset{\text{(\ref{leg_dependent_mass_rule})}}{\longmapsto}}-\frac{\gamma_c(\coupling)}{16}A_{n,\higgsRegName}^{(1)}\!+\!\frac{\widetilde{\cal G}_0(\coupling)}{2}\Big[n\log(\eps)+\frac{1}{2}\log(w_1\cdots w_n)\Big]\\
&+n\widetilde{f}(\coupling)+\widetilde{C}(\coupling)+R_n(\coupling)
\end{split}}
where $\gamma_c(\coupling)$ is the (scheme-independent) \emph{cusp anomalous dimension} \cite{Sterman:2002qn,Beisert:2006ez}
\eq{\gamma_c(\coupling)\equivL\sum_{\ell=1}^{\infty}\coupling^\ell\gamma_c^{(\ell)}=4\coupling-4\Zeta{2}\coupling^2+22\Zeta{4}\coupling^3-\Big(24\Zeta{2}^3+4\Zeta{3}^2+2\Zeta{2}\Zeta{4}+\Zeta{6}\Big)\coupling^4+\mathcal{O}(\coupling^5)\,,}
 $\widetilde{\cal G}_0(\coupling), \widetilde{f}(\coupling), \widetilde{C}(\coupling)$ are scheme-dependent functions of the coupling and $R_n(\coupling)$ is the \emph{remainder function} \cite{Bern:2008ap}. In the Higgs regularization scheme these functions were determined by \cite{Alday:2009zm,Henn:2010ir} to be
\eq{\widetilde{\cal G}_0(\coupling)=-\Zeta{3}\coupling^2+\mathcal{O}(\coupling^3),\quad\widetilde{f}(\coupling)=\frac{1}{2}\Zeta{4}\coupling^2+\mathcal{O}(\coupling^3),\quad\widetilde{C}(\coupling)=-\frac{5}{4}\Zeta{4}\coupling^2+\mathcal{O}(\coupling^3)\,,}
at two-loop order. (See e.g.\ \mbox{\cite{Correa:2012nk,Henn:2013wfa}} for more recent, higher-order results.)

With this comparison in mind, let us now return to the main purpose of this work and describe the form the logarithm takes for the  conformal regularization scheme.

\vspace{-0pt}\subsection{Conformally-Regulated (Logarithms of) MHV Amplitudes}\label{subsec:conformally_regulated_amps}\vspace{-0pt}

Using the conformal regulator described in \mbox{\cite{Bourjaily:2013mma}} the divergences of one-loop amplitudes take a form strikingly similar to that of (\ref{conformal_like_version_of_higgs_one_loop}). In this scheme, the $n$-point MHV amplitude is given by\footnote{We have added a factor of $1/2$ relative to \mbox{\cite{Bourjaily:2013mma}} to match conventions for the coupling $\coupling$.}
\eq{A_{n,\dciRegName}^{(1)}\equivR -\frac{1}{2} \Bigg[n\log^2\!(\eps)+\log(\eps)\log(w_1\cdots w_n)+n\Zeta{2}+F_{n,\dciRegName}^{(1)}\Bigg]+\mathcal{O}(\eps) \,,\label{conformally_regulated_one_loop}}
where the cross-ratios $w_a$ are the same as those defined in (\ref{w_cross_ratios_defined}) and 
\eq{F_{n,\dciRegName}^{(1)}=\Bigg[\sum_{{\color{hred}b}=4}^{\lfloor n/2\rfloor+1}\!\!\Li{2}{1-u_{{\color{hblue}1},{\color{hred}b}}}+\frac{1}{2}\log(u_{{\color{hblue}1},{\color{hred}b}})\log(v_{{\color{hblue}1},{\color{hred}b}})\Bigg]+\cyclicPrime}
where the cross-ratios $u_{{\color{hblue}a},{\color{hred}b}}$ and $v_{{\color{hblue}a},{\color{hred}b}}$ are given by 
\eq{u_{{\color{hblue}a},{\color{hred}b}}\equivR\frac{\x{{\color{hblue}a+1}}{{\color{hred}b}}\x{{\color{hred}b+1}}{{\color{hblue}a}}}{\x{{\color{hblue}a+1}}{{\color{hred}b+1}}\x{{\color{hred}b}}{{\color{hblue}a}}}\,,\quad v_{{\color{hblue}a},{\color{hred}b}}\equivR\frac{\x{{\color{hblue}a-1}}{{\color{hblue}a+1}}\x{{\color{hblue}a}}{{\color{hblue}a+2}}\x{{\color{hred}b-1}}{{\color{hred}b+1}}\x{{\color{hred}b}}{{\color{hred}b+2}}}{\x{{\color{hblue}a-1}}{{\color{hblue}a+2}}\x{{\color{hblue}a}}{{\color{hred}b}}\x{{\color{hred}b-1}}{{\color{hred}b+2}}\x{{\color{hred}b+1}}{{\color{hblue}a+1}}}\,.}

In terms of the regulated amplitude at one loop (\ref{conformally_regulated_one_loop}), it was suggested in \mbox{\cite{Bourjaily:2019jrk}} that the conformally regulated logarithm (\ref{generic_log_expansion}) would take the form 
\eq{\begin{split}\hspace{-20pt}\log(A_{n,\dciRegName})\equivL&-\frac{\gamma_c(\coupling)}{8}A_{n,\dciRegName}^{(1)}\!+\!\frac{B_{\delta}(\coupling)}{2}\Big[n\log(\eps)+n+\frac{1}{2}\log(w_1\cdots w_n)\Big]\\
&+n\widehat{f}(\coupling)+\widehat{C}(\coupling)+R_n(\coupling)\hspace{-20pt}
\end{split}}
where $B_\delta(\coupling)\equivR3\Zeta{3}\coupling^2+\mathcal{O}(\coupling^3)$ is the so-called \emph{virtual} anomalous dimension~\cite{Kotikov:2004er,Freyhult:2007pz}, and the functions $\widehat{f}(\coupling)$ and $\widehat{C}(\coupling)$ are analogous to $\widetilde{f}(\coupling)$ and $\widetilde{C}(\coupling)$---which could not be disentangled from each other knowing the logarithm for six particles alone. 

In \mbox{\cite{Bourjaily:2019jrk}}, the six-point logarithm was shown to take the form\footnote{\emph{nota bene}: for six particles, $(w_1\cdots w_6)=(w_1w_2w_3)^2$, with $w_i$ more familiarly denoted $\{u,v,w\}$.}
\eq{\log(A_{6,\dciRegName})^{(2)}=-\Zeta{2}A_{6,\dciRegName}^{(1)}+\frac{3}{2}\Zeta{3}\Big[6\log(\eps)+6+\frac{1}{2}\log(w_1\cdots w_6)\Big]-\frac{49\,\pi^4}{720}+R_6^{(2)}\,;}
and for five particles, starting from representation given in~(\ref{5pt_log_integrand}), it is not hard to show that\footnote{\emph{nota bene}: for five particles, $w_a=1$ for all $a$ and $R_5^{(\ell)}=0$ for all $\ell$.}
\eq{\log(A_{5,\dciRegName})^{(2)}=-\Zeta{2}A_{5,\dciRegName}^{(1)}+\frac{3}{2}\Zeta{3}\Big[5\log(\eps)+5+\log(w_1\cdots w_5)\Big]-\frac{17\,\pi^4}{288}+R_5^{(2)}\,.}
Combining this with our new result for seven particles, 
\eq{\log(A_{7,\dciRegName})^{(2)}=-\Zeta{2}A_{7,\dciRegName}^{(1)}+\frac{3}{2}\Zeta{3}\Big[7\log(\eps)+7+\frac{1}{2}\log(w_1\cdots w_7)\Big]-\frac{37\,\pi^4}{480}+R_7^{(2)}\,,\label{seven_point_log}}
allows us to conclude that, in the conformal regularization scheme, 
\eq{\widehat{f}(\coupling)=-\frac{1}{2}\Big(\Zeta{4}+\frac{1}{4}\Zeta{2}^2\Big)\coupling^2+\mathcal{O}(\coupling^3)\,,\quad\widehat{C}(\coupling)=-\frac{1}{2}\Zeta{2}^2\coupling^2+\mathcal{O}(\coupling^3)\,.}

Although already mentioned in the introduction, it is worth pausing to note that, in the representation of the logarithm (\ref{seven_point_log}), the remainder function $R_7^{(2)}$ numerically matches the analytic expression derived in \mbox{\cite{Golden:2014xqf}} from the symbol (from \mbox{\cite{CaronHuot:2011ky}}). 

\vspace{-0pt}\subsection{Symbology and the Alphabets of Individual Integral Contributions}\label{subsec:symbol_letters_discussion}\vspace{-0pt}

Interestingly, almost all of the seed integrals we compute contain symbol letters that are not present in the full remainder function.
The integral $\seedA$ is the only exception: it in fact requires only the ordinary hexagon-function symbol alphabet. 
However, each of the other integrals involve spurious (but rational) symbol letters. Specifically, each of $\{\seedB,\seedC,\seedD\}$ involve two `new' letters relative to the remainder function, and $\seedE$ involves nine additional letters (after all the simplifications described in subsection \ref{subsec:processing_integrals}). In cyclic sum, however, all these additional letters cancel---and quite nontrivially. For example, among these contributions only the \emph{entire} cyclic sum of \mbox{$\big(\seedB\pl\seedC\pl\seedD\pl\seedE\big)$} is free of `spurious' letters relative to the 42 letter alphabet expected for heptagon functions \cite{Golden:2014xqf} (see also \cite{Golden:2013lha,Golden:2013xva,Drummond:2014ffa,Golden:2014pua,Golden:2014xqa,Dixon:2016nkn,Drummond:2017ssj,Drummond:2018caf}). For the sake of those readers interested in more details, we have provided the additional symbol letters that arise for the cyclic seed integrals in the ancillary files to this work.

\vspace{-0pt}\section{Discussion}\label{sec:conclusions}\vspace{-0pt}

In this paper, we have computed the logarithm of the two-loop MHV amplitude at seven points in planar, maximally supersymmetric ($\mathcal{N}\!=\!4$) super Yang-Mills theory directly from a local integrand representation. In doing so, we have shown that carefully preserving the symmetries of the theory makes computations dramatically easier, even when using otherwise traditional methods. However, these methods are still not optimal: as we have seen, issues of linear reducibility make some of the integrals we find unsuitable for expansion into a fibration basis (by known methods), resulting in a sometimes unnecessarily-spurious symbol alphabet. It would be interesting to see whether other common methods (for example, differential equations, or integration-by-parts reduction) can simplify this calculation further.

In using the dual conformal regularization of \cite{Bourjaily:2019jrk}, we have checked the conjectures for the scheme dependence of the logarithm of the amplitude put forward in that paper. It would be interesting to check these conjectures at higher loop orders, and more generally, to understand in detail the relationship between the conformal regulator and the Higgs regulator.

\vspace{\fill}\vspace{-4pt}
\section*{Acknowledgements}%
\vspace{-4pt}
This work was supported in part by the Danish National Research Foundation (DNRF91), an ERC Starting Grant \mbox{(No.\ 757978)}, a grant from the Villum Fonden \mbox{(No.\ 15369)}, and the European Union's Horizon 2020 research and innovation program under grant agreement \mbox{No.\ 793151} (MvH). We are grateful for the hospitality of the Harvard Center for Mathematical Sciences and Applications, and JLB and MvH would like to thank the Aspen Center for Physics, which is supported by National Science Foundation grant PHY-1607611.

\providecommand{\href}[2]{#2}\begingroup\raggedright\endgroup


\begin{thebibliography}{10}

\bibitem{ArkaniHamed:2009dn}
N.~Arkani-Hamed, F.~Cachazo, C.~Cheung, and J.~Kaplan, ``{A Duality For The
  $S$-Matrix},'' \href{http://dx.doi.org/10.1007/JHEP03(2010)020}{{\em JHEP}
  {\bf 1003} (2010)  020},
\href{http://arxiv.org/abs/0907.5418}{{ arXiv:0907.5418 [hep-th]}}.

\bibitem{ArkaniHamed:2009dg}
N.~Arkani-Hamed, J.~Bourjaily, F.~Cachazo, and J.~Trnka, ``{Unification of
  Residues and Grassmannian Dualities},''
  \href{http://dx.doi.org/10.1007/JHEP01(2011)049}{{\em JHEP} {\bf 1101} (2011)
   049},
\href{http://arxiv.org/abs/0912.4912}{{ arXiv:0912.4912 [hep-th]}}.

\bibitem{ArkaniHamed:2012nw}
N.~Arkani-Hamed, J.~L. Bourjaily, F.~Cachazo, A.~B. Goncharov, A.~Postnikov,
  and J.~Trnka, ``{Scattering Amplitudes and the Positive Grassmannian},''
\href{http://arxiv.org/abs/1212.5605}{{ arXiv:1212.5605 [hep-th]}}.

\bibitem{Arkani-Hamed:2013jha}
N.~Arkani-Hamed and J.~Trnka, ``{The Amplituhedron},''
  \href{http://dx.doi.org/10.1007/JHEP10(2014)030}{{\em JHEP} {\bf 1410} (2014)
   30},
\href{http://arxiv.org/abs/1312.2007}{{ arXiv:1312.2007 [hep-th]}}.

\bibitem{Frye:2018xjj}
C.~Frye, H.~Hannesdottir, N.~Paul, M.~D. Schwartz, and K.~Yan, ``{Infrared
  Finiteness and Forward Scattering},''
  \href{http://dx.doi.org/10.1103/PhysRevD.99.056015}{{\em Phys. Rev.} {\bf
  D99} (2019) no. 5, 056015},
\href{http://arxiv.org/abs/1810.10022}{{ arXiv:1810.10022 [hep-ph]}}.

\bibitem{Hannesdottir:2019rqq}
H.~Hannesdottir and M.~D. Schwartz, ``{A Finite $S$-Matrix},''
\href{http://arxiv.org/abs/1906.03271}{{ arXiv:1906.03271 [hep-th]}}.

\bibitem{ArkaniHamed:2010kv}
N.~Arkani-Hamed, J.~L. Bourjaily, F.~Cachazo, S.~Caron-Huot, and J.~Trnka,
  ``{The All-Loop Integrand For Scattering Amplitudes in Planar
  $\mathcal{N}\!=\!4$ SYM},''
  \href{http://dx.doi.org/10.1007/JHEP01(2011)041}{{\em JHEP} {\bf 1101} (2011)
   041},
\href{http://arxiv.org/abs/1008.2958}{{ arXiv:1008.2958 [hep-th]}}.

\bibitem{Benincasa:2015zna}
P.~Benincasa, ``{On-Shell Diagrammatics and the Perturbative Structure of
  Planar Gauge Theories},''
\href{http://arxiv.org/abs/1510.03642}{{ arXiv:1510.03642 [hep-th]}}.

\bibitem{Bourjaily:2011hi}
J.~L. Bourjaily, A.~DiRe, A.~Shaikh, M.~Spradlin, and A.~Volovich, ``{The
  Soft-Collinear Bootstrap: $\mathcal{N}\!=\!4$ Yang-Mills Amplitudes at Six
  and Seven Loops},'' \href{http://dx.doi.org/10.1007/JHEP03(2012)032}{{\em
  JHEP} {\bf 1203} (2012)  032},
\href{http://arxiv.org/abs/1112.6432}{{ arXiv:1112.6432 [hep-th]}}.

\bibitem{Bourjaily:2015bpz}
J.~L. Bourjaily, P.~Heslop, and V.-V. Tran, ``{Perturbation Theory at Eight
  Loops: Novel Structures and the Breakdown of Manifest Conformality in
  $\mathcal{N}\!=\!4$ Supersymmetric Yang-Mills Theory},''
  \href{http://dx.doi.org/10.1103/PhysRevLett.116.191602}{{\em Phys. Rev.
  Lett.} {\bf 116} (2016) no. 19, 191602},
\href{http://arxiv.org/abs/1512.07912}{{ arXiv:1512.07912 [hep-th]}}.

\bibitem{Bourjaily:2016evz}
J.~L. Bourjaily, P.~Heslop, and V.-V. Tran, ``{Amplitudes and Correlators to
  Ten Loops Using Simple, Graphical Bootstraps},''
  \href{http://dx.doi.org/10.1007/JHEP11(2016)125}{{\em JHEP} {\bf 11} (2016)
  125},
\href{http://arxiv.org/abs/1609.00007}{{ arXiv:1609.00007 [hep-th]}}.

\bibitem{Baadsgaard:2015twa}
C.~Baadsgaard, N.~E.~J. Bjerrum-Bohr, J.~L. Bourjaily, S.~Caron-Huot, P.~H.
  Damgaard, and B.~Feng, ``{New Representations of the Perturbative
  $S$-Matrix},'' \href{http://dx.doi.org/10.1103/PhysRevLett.116.061601}{{\em
  Phys. Rev. Lett.} {\bf 116} (2016) no. 6, 061601},
\href{http://arxiv.org/abs/1509.02169}{{ arXiv:1509.02169 [hep-th]}}.

\bibitem{Bern:1994zx}
Z.~Bern, L.~J. Dixon, D.~C. Dunbar, and D.~A. Kosower, ``{One-Loop $n$-Point
  Gauge Theory Amplitudes, Unitarity and Collinear Limits},''
  \href{http://dx.doi.org/10.1016/0550-3213(94)90179-1}{{\em Nucl. Phys.} {\bf
  B425} (1994)  217--260},
\href{http://arxiv.org/abs/hep-ph/9403226}{{ arXiv:hep-ph/9403226}}.

\bibitem{Bern:1994cg}
Z.~Bern, L.~J. Dixon, D.~C. Dunbar, and D.~A. Kosower, ``{Fusing Gauge Theory
  Tree Amplitudes into Loop Amplitudes},''
  \href{http://dx.doi.org/10.1016/0550-3213(94)00488-Z}{{\em Nucl. Phys.} {\bf
  B435} (1995)  59--101},
\href{http://arxiv.org/abs/hep-ph/9409265}{{ arXiv:hep-ph/9409265}}.

\bibitem{Britto:2004nc}
R.~Britto, F.~Cachazo, and B.~Feng, ``{Generalized Unitarity and One-Loop
  Amplitudes in $\mathcal{N}\!=\!4$ Super-Yang-Mills},''
  \href{http://dx.doi.org/10.1016/j.nuclphysb.2005.07.014}{{\em Nucl. Phys.}
  {\bf B725} (2005)  275--305},
\href{http://arxiv.org/abs/hep-th/0412103}{{ arXiv:hep-th/0412103}}.

\bibitem{Bern:2006ew}
Z.~Bern, M.~Czakon, L.~J. Dixon, D.~A. Kosower, and V.~A. Smirnov, ``{The
  Four-Loop Planar Amplitude and Cusp Anomalous Dimension in Maximally
  Supersymmetric Yang-Mills Theory},''
  \href{http://dx.doi.org/10.1103/PhysRevD.75.085010}{{\em Phys. Rev.} {\bf
  D75} (2007)  085010},
\href{http://arxiv.org/abs/hep-th/0610248}{{ arXiv:hep-th/0610248 [hep-th]}}.

\bibitem{Anastasiou:2006jv}
C.~Anastasiou, R.~Britto, B.~Feng, Z.~Kunszt, and P.~Mastrolia,
  ``{$D$-Dimensional Unitarity Cut Method},''
  \href{http://dx.doi.org/10.1016/j.physletb.2006.12.022}{{\em Phys. Lett.}
  {\bf B645} (2007)  213--216},
\href{http://arxiv.org/abs/hep-ph/0609191}{{ arXiv:hep-ph/0609191 [hep-ph]}}.

\bibitem{Bern:2007ct}
Z.~Bern, J.~Carrasco, H.~Johansson, and D.~Kosower, ``{Maximally Supersymmetric
  Planar Yang-Mills Amplitudes at Five Loops},''
  \href{http://dx.doi.org/10.1103/PhysRevD.76.125020}{{\em Phys. Rev.} {\bf
  D76} (2007)  125020},
\href{http://arxiv.org/abs/0705.1864}{{ arXiv:0705.1864 [hep-th]}}.

\bibitem{Cachazo:2008vp}
F.~Cachazo, ``{Sharpening The Leading Singularity},''
\href{http://arxiv.org/abs/0803.1988}{{ arXiv:0803.1988 [hep-th]}}.

\bibitem{Bern:2008ap}
Z.~Bern {\em et al.}, ``{The Two-Loop Six-Gluon MHV Amplitude in Maximally
  Supersymmetric Yang-Mills Theory},''
  \href{http://dx.doi.org/10.1103/PhysRevD.78.045007}{{\em Phys. Rev.} {\bf
  D78} (2008)  045007},
\href{http://arxiv.org/abs/0803.1465}{{ arXiv:0803.1465 [hep-th]}}.

\bibitem{Berger:2008sj}
C.~F. Berger {\em et al.}, ``{An Automated Implementation of On-Shell Methods
  for One-Loop Amplitudes},''
  \href{http://dx.doi.org/10.1103/PhysRevD.78.036003}{{\em Phys. Rev.} {\bf
  D78} (2008)  036003},
\href{http://arxiv.org/abs/0803.4180}{{ arXiv:0803.4180 [hep-ph]}}.

\bibitem{Abreu:2017xsl}
S.~Abreu, F.~Febres~Cordero, H.~Ita, M.~Jaquier, B.~Page, and M.~Zeng,
  ``{Two-Loop Four-Gluon Amplitudes with the Numerical Unitarity Method},''
\href{http://arxiv.org/abs/1703.05273}{{ arXiv:1703.05273 [hep-ph]}}.

\bibitem{ArkaniHamed:2010gh}
N.~Arkani-Hamed, J.~L. Bourjaily, F.~Cachazo, and J.~Trnka, ``{Local Integrals
  for Planar Scattering Amplitudes},''
  \href{http://dx.doi.org/10.1007/JHEP06(2012)125}{{\em JHEP} {\bf 1206} (2012)
   125},
\href{http://arxiv.org/abs/1012.6032}{{ arXiv:1012.6032 [hep-th]}}.

\bibitem{Bourjaily:2013mma}
J.~L. Bourjaily, S.~Caron-Huot, and J.~Trnka, ``{Dual-Conformal Regularization
  of Infrared Loop Divergences and the {\it Chiral} Box Expansion},''
  \href{http://dx.doi.org/10.1007/JHEP01(2015)001}{{\em JHEP} {\bf 1501} (2015)
   001},
\href{http://arxiv.org/abs/1303.4734}{{ arXiv:1303.4734 [hep-th]}}.

\bibitem{Bourjaily:2015jna}
J.~L. Bourjaily and J.~Trnka, ``{Local Integrand Representations of All
  Two-Loop Amplitudes in Planar SYM},''
  \href{http://dx.doi.org/10.1007/JHEP08(2015)119}{{\em JHEP} {\bf 08} (2015)
  119},
\href{http://arxiv.org/abs/1505.05886}{{ arXiv:1505.05886 [hep-th]}}.

\bibitem{Bourjaily:2017wjl}
J.~L. Bourjaily, E.~Herrmann, and J.~Trnka, ``{Prescriptive Unitarity},''
  \href{http://dx.doi.org/10.1007/JHEP06(2017)059}{{\em JHEP} {\bf 06} (2017)
  059},
\href{http://arxiv.org/abs/1704.05460}{{ arXiv:1704.05460 [hep-th]}}.

\bibitem{Bourjaily:2018omh}
J.~L. Bourjaily, E.~Herrmann, and J.~Trnka, ``{Amplitudes at Infinity},''
  \href{http://dx.doi.org/10.1103/PhysRevD.99.066006}{{\em Phys. Rev.} {\bf
  D99} (2019) no. 6, 066006},
\href{http://arxiv.org/abs/1812.11185}{{ arXiv:1812.11185 [hep-th]}}.

\bibitem{Bourjaily:2019iqr}
J.~L. Bourjaily, E.~Herrmann, C.~Langer, A.~J. McLeod, and J.~Trnka,
  ``{Prescriptive Unitarity for Non-Planar Six-Particle Amplitudes at Two
  Loops},''
\href{http://arxiv.org/abs/1909.09131}{{ arXiv:1909.09131 [hep-th]}}.

\bibitem{Bourjaily:2019gqu}
J.~L. Bourjaily, E.~Herrmann, C.~Langer, A.~J. McLeod, and J.~Trnka,
  ``{All-Multiplicity Non-Planar MHV Amplitudes in sYM at Two Loops},''
\href{http://arxiv.org/abs/1911.09106}{{ arXiv:1911.09106 [hep-th]}}.

\bibitem{Mandelstam:1982cb}
S.~Mandelstam, ``{Light Cone Superspace and the Ultraviolet Finiteness of the
  $\mathcal{N}\!=\!4$ Model},''
\href{http://dx.doi.org/10.1016/0550-3213(83)90179-7}{{\em Nucl. Phys.} {\bf
  B213} (1983)  149--168}.

\bibitem{Brink:1982wv}
L.~Brink, O.~Lindgren, and B.~E.~W. Nilsson, ``{The Ultraviolet Finiteness of
  the $\mathcal{N}\!=\!4$ Yang-Mills Theory},''
\href{http://dx.doi.org/10.1016/0370-2693(83)91210-8}{{\em Phys. Lett.} {\bf
  123B} (1983)  323--328}.

\bibitem{Howe:1983sr}
P.~S. Howe, K.~S. Stelle, and P.~K. Townsend, ``{Miraculous Ultraviolet
  Cancellations in Supersymmetry Made Manifest},''
\href{http://dx.doi.org/10.1016/0550-3213(84)90528-5}{{\em Nucl. Phys.} {\bf
  B236} (1984)  125}.

\bibitem{Drummond:2006rz}
J.~Drummond, J.~Henn, V.~Smirnov, and E.~Sokatchev, ``{Magic Identities for
  Conformal Four-Point Integrals},''
  \href{http://dx.doi.org/10.1088/1126-6708/2007/01/064}{{\em JHEP} {\bf 0701}
  (2007)  064},
\href{http://arxiv.org/abs/hep-th/0607160}{{ arXiv:hep-th/0607160}}.

\bibitem{Alday:2007hr}
L.~F. Alday and J.~M. Maldacena, ``{Gluon Scattering Amplitudes at Strong
  Coupling},'' \href{http://dx.doi.org/10.1088/1126-6708/2007/06/064}{{\em
  JHEP} {\bf 06} (2007)  064},
\href{http://arxiv.org/abs/0705.0303}{{ arXiv:0705.0303 [hep-th]}}.

\bibitem{Drummond:2008vq}
J.~Drummond, J.~Henn, G.~Korchemsky, and E.~Sokatchev, ``{Dual Superconformal
  Symmetry of Scattering Amplitudes in $\mathcal{N}\!=\!4$ super Yang-Mills
  Theory},'' \href{http://dx.doi.org/10.1016/j.nuclphysb.2009.11.022}{{\em
  Nucl. Phys.} {\bf B828} (2010)  317--374},
\href{http://arxiv.org/abs/0807.1095}{{ arXiv:0807.1095 [hep-th]}}.

\bibitem{DelDuca:2010zg}
V.~Del~Duca, C.~Duhr, and V.~A. Smirnov, ``{The Two-Loop Hexagon Wilson Loop in
  $\mathcal{N}\!=\!4$ SYM},''
  \href{http://dx.doi.org/10.1007/JHEP05(2010)084}{{\em JHEP} {\bf 05} (2010)
  084},
\href{http://arxiv.org/abs/1003.1702}{{ arXiv:1003.1702 [hep-th]}}.

\bibitem{Goncharov:2010jf}
A.~B. Goncharov, M.~Spradlin, C.~Vergu, and A.~Volovich, ``{Classical
  Polylogarithms for Amplitudes and Wilson Loops},''
  \href{http://dx.doi.org/10.1103/PhysRevLett.105.151605}{{\em Phys. Rev.
  Lett.} {\bf 105} (2010)  151605},
\href{http://arxiv.org/abs/1006.5703}{{ arXiv:1006.5703 [hep-th]}}.

\bibitem{Dixon:2011nj}
L.~J. Dixon, J.~M. Drummond, and J.~M. Henn, ``{Analytic Result for the
  Two-Loop Six-Point NMHV Amplitude in $\mathcal{N}\!=\!4$ Super Yang-Mills
  Theory},'' \href{http://dx.doi.org/10.1007/JHEP01(2012)024}{{\em JHEP} {\bf
  1201} (2012)  024},
\href{http://arxiv.org/abs/1111.1704}{{ arXiv:1111.1704 [hep-th]}}.

\bibitem{Dixon:2013eka}
L.~J. Dixon, J.~M. Drummond, M.~von Hippel, and J.~Pennington, ``{Hexagon
  Functions and the Three-Loop Remainder Function},''
  \href{http://dx.doi.org/10.1007/JHEP12(2013)049}{{\em JHEP} {\bf 1312} (2013)
   049},
\href{http://arxiv.org/abs/1308.2276}{{ arXiv:1308.2276 [hep-th]}}.

\bibitem{Dixon:2014voa}
L.~J. Dixon, J.~M. Drummond, C.~Duhr, and J.~Pennington, ``{The Four-Loop
  Remainder Function and Multi-Regge Behavior at NNLLA in Planar
  $\mathcal{N}\!=\!4$ Super-Yang-Mills Theory},''
  \href{http://dx.doi.org/10.1007/JHEP06(2014)116}{{\em JHEP} {\bf 1406} (2014)
   116},
\href{http://arxiv.org/abs/1402.3300}{{ arXiv:1402.3300 [hep-th]}}.

\bibitem{Dixon:2014iba}
L.~J. Dixon and M.~von Hippel, ``{Bootstrapping an NMHV Amplitude Through Three
  Loops},'' \href{http://dx.doi.org/10.1007/JHEP10(2014)065}{{\em JHEP} {\bf
  1410} (2014)  65},
\href{http://arxiv.org/abs/1408.1505}{{ arXiv:1408.1505 [hep-th]}}.

\bibitem{Dixon:2015iva}
L.~J. Dixon, M.~von Hippel, and A.~J. McLeod, ``{The Four-Loop Six-Gluon NMHV
  Ratio Function},'' \href{http://dx.doi.org/10.1007/JHEP01(2016)053}{{\em
  JHEP} {\bf 01} (2016)  053},
\href{http://arxiv.org/abs/1509.08127}{{ arXiv:1509.08127 [hep-th]}}.

\bibitem{Dixon:2016apl}
L.~J. Dixon, M.~von Hippel, A.~J. McLeod, and J.~Trnka, ``{Multi-Loop
  Positivity of the Planar $\mathcal{N}\!=\!4$ SYM Six-Point Amplitude},''
  \href{http://dx.doi.org/10.1007/JHEP02(2017)112}{{\em JHEP} {\bf 02} (2017)
  112},
\href{http://arxiv.org/abs/1611.08325}{{ arXiv:1611.08325 [hep-th]}}.

\bibitem{Caron-Huot:2016owq}
S.~Caron-Huot, L.~J. Dixon, A.~McLeod, and M.~von Hippel, ``{Bootstrapping a
  Five-Loop Amplitude Using Steinmann Relations},''
  \href{http://dx.doi.org/10.1103/PhysRevLett.117.241601}{{\em Phys. Rev.
  Lett.} {\bf 117} (2016) no. 24, 241601},
\href{http://arxiv.org/abs/1609.00669}{{ arXiv:1609.00669 [hep-th]}}.

\bibitem{Caron-Huot:2019vjl}
S.~Caron-Huot, L.~J. Dixon, F.~Dulat, M.~von Hippel, A.~J. McLeod, and
  G.~Papathanasiou, ``{Six-Gluon Amplitudes in Planar $\mathcal{N}\!=\!4 $
  super-Yang-Mills Theory at Six and Seven Loops},''
  \href{http://dx.doi.org/10.1007/JHEP08(2019)016}{{\em JHEP} {\bf 08} (2019)
  016},
\href{http://arxiv.org/abs/1903.10890}{{ arXiv:1903.10890 [hep-th]}}.

\bibitem{Caron-Huot:2019bsq}
S.~Caron-Huot, L.~J. Dixon, F.~Dulat, M.~Von~Hippel, A.~J. McLeod, and
  G.~Papathanasiou, ``{The Cosmic Galois Group and Extended Steinmann Relations
  for Planar $\mathcal{N} = 4$ SYM Amplitudes},''
\href{http://arxiv.org/abs/1906.07116}{{ arXiv:1906.07116 [hep-th]}}.

\bibitem{Drummond:2014ffa}
J.~M. Drummond, G.~Papathanasiou, and M.~Spradlin, ``{A Symbol of Uniqueness:
  The Cluster Bootstrap for the 3-Loop MHV Heptagon},''
  \href{http://dx.doi.org/10.1007/JHEP03(2015)072}{{\em JHEP} {\bf 03} (2015)
  072},
\href{http://arxiv.org/abs/1412.3763}{{ arXiv:1412.3763 [hep-th]}}.

\bibitem{Dixon:2016nkn}
L.~J. Dixon, J.~Drummond, T.~Harrington, A.~J. McLeod, G.~Papathanasiou, and
  M.~Spradlin, ``{Heptagons from the Steinmann Cluster Bootstrap},''
  \href{http://dx.doi.org/10.1007/JHEP02(2017)137}{{\em JHEP} {\bf 02} (2017)
  137},
\href{http://arxiv.org/abs/1612.08976}{{ arXiv:1612.08976 [hep-th]}}.

\bibitem{Drummond:2018caf}
J.~Drummond, J.~Foster, {\"O}.~G{\"u}rdo\v{g}an, and G.~Papathanasiou,
  ``{Cluster Adjacency and the Four-Loop NMHV Heptagon},''
  \href{http://dx.doi.org/10.1007/JHEP03(2019)087}{{\em JHEP} {\bf 03} (2019)
  087},
\href{http://arxiv.org/abs/1812.04640}{{ arXiv:1812.04640 [hep-th]}}.

\bibitem{Abreu:2018aqd}
S.~Abreu, L.~J. Dixon, E.~Herrmann, B.~Page, and M.~Zeng, ``{The Two-Loop
  Five-Point Amplitude in $\mathcal{N}\!=\!4$ super-Yang-Mills Theory},''
  \href{http://dx.doi.org/10.1103/PhysRevLett.122.121603}{{\em Phys. Rev.
  Lett.} {\bf 122} (2019) no. 12, 121603},
\href{http://arxiv.org/abs/1812.08941}{{ arXiv:1812.08941 [hep-th]}}.

\bibitem{Chicherin:2018yne}
D.~Chicherin, T.~Gehrmann, J.~M. Henn, P.~Wasser, Y.~Zhang, and S.~Zoia,
  ``{Analytic Result for a Two-Loop Five-Particle Amplitude},''
  \href{http://dx.doi.org/10.1103/PhysRevLett.122.121602}{{\em Phys. Rev.
  Lett.} {\bf 122} (2019) no. 12, 121602},
\href{http://arxiv.org/abs/1812.11057}{{ arXiv:1812.11057 [hep-th]}}.

\bibitem{Gehrmann:2015bfy}
T.~Gehrmann, J.~M. Henn, and N.~A. Lo~Presti, ``{Analytic Form of the Two-Loop
  Planar Five-Gluon All-Plus-Helicity Amplitude in QCD},''
  \href{http://dx.doi.org/10.1103/PhysRevLett.116.189903,
  10.1103/PhysRevLett.116.062001}{{\em Phys. Rev. Lett.} {\bf 116} (2016) no.
  6, 062001}, \href{http://arxiv.org/abs/1511.05409}{{ arXiv:1511.05409
  [hep-ph]}}.
[Erratum: Phys. Rev. Lett.116,no.18,189903(2016)].

\bibitem{Badger:2017jhb}
S.~Badger, C.~Br{\o}nnum-Hansen, H.~B. Hartanto, and T.~Peraro, ``{First Look
  at Two-Loop Five-Gluon Scattering in QCD},''
  \href{http://dx.doi.org/10.1103/PhysRevLett.120.092001}{{\em Phys. Rev.
  Lett.} {\bf 120} (2018) no. 9, 092001},
\href{http://arxiv.org/abs/1712.02229}{{ arXiv:1712.02229 [hep-ph]}}.

\bibitem{Abreu:2017hqn}
S.~Abreu, F.~Febres~Cordero, H.~Ita, B.~Page, and M.~Zeng, ``{Planar Two-Loop
  Five-Gluon Amplitudes from Numerical Unitarity},''
  \href{http://dx.doi.org/10.1103/PhysRevD.97.116014}{{\em Phys. Rev.} {\bf
  D97} (2018) no. 11, 116014},
\href{http://arxiv.org/abs/1712.03946}{{ arXiv:1712.03946 [hep-ph]}}.

\bibitem{Abreu:2018jgq}
S.~Abreu, F.~Febres~Cordero, H.~Ita, B.~Page, and V.~Sotnikov, ``{Planar
  Two-Loop Five-Parton Amplitudes from Numerical Unitarity},''
  \href{http://dx.doi.org/10.1007/JHEP11(2018)116}{{\em JHEP} {\bf 11} (2018)
  116},
\href{http://arxiv.org/abs/1809.09067}{{ arXiv:1809.09067 [hep-ph]}}.

\bibitem{Badger:2018enw}
S.~Badger, C.~Br{\o}nnum-Hansen, H.~B. Hartanto, and T.~Peraro, ``{Analytic
  Helicity Amplitudes for Two-Loop Five-Gluon Scattering: the Single-Minus
  Case},'' \href{http://dx.doi.org/10.1007/JHEP01(2019)186}{{\em JHEP} {\bf 01}
  (2019)  186},
\href{http://arxiv.org/abs/1811.11699}{{ arXiv:1811.11699 [hep-ph]}}.

\bibitem{Abreu:2018zmy}
S.~Abreu, J.~Dormans, F.~Febres~Cordero, H.~Ita, and B.~Page, ``{Analytic Form
  of Planar Two-Loop Five-Gluon Scattering Amplitudes in QCD},''
  \href{http://dx.doi.org/10.1103/PhysRevLett.122.082002}{{\em Phys. Rev.
  Lett.} {\bf 122} (2019) no. 8, 082002},
\href{http://arxiv.org/abs/1812.04586}{{ arXiv:1812.04586 [hep-ph]}}.

\bibitem{Chicherin:2018old}
D.~Chicherin, T.~Gehrmann, J.~M. Henn, P.~Wasser, Y.~Zhang, and S.~Zoia, ``{All
  Master Integrals for Three-Jet Production at Next-to-Next-to-Leading
  Order},'' \href{http://dx.doi.org/10.1103/PhysRevLett.123.041603}{{\em Phys.
  Rev. Lett.} {\bf 123} (2019) no. 4, 041603},
\href{http://arxiv.org/abs/1812.11160}{{ arXiv:1812.11160 [hep-ph]}}.

\bibitem{Chicherin:2019xeg}
D.~Chicherin, T.~Gehrmann, J.~M. Henn, P.~Wasser, Y.~Zhang, and S.~Zoia, ``{The
  Two-Loop Five-Particle Amplitude in $\mathcal{N}\!=\!8$ Supergravity},''
  \href{http://dx.doi.org/10.1007/JHEP03(2019)115}{{\em JHEP} {\bf 03} (2019)
  115},
\href{http://arxiv.org/abs/1901.05932}{{ arXiv:1901.05932 [hep-th]}}.

\bibitem{Abreu:2019rpt}
S.~Abreu, L.~J. Dixon, E.~Herrmann, B.~Page, and M.~Zeng, ``{The Two-Loop
  Five-Point Amplitude in $ \mathcal{N}\!=\!8$ Supergravity},''
  \href{http://dx.doi.org/10.1007/JHEP03(2019)123}{{\em JHEP} {\bf 03} (2019)
  123},
\href{http://arxiv.org/abs/1901.08563}{{ arXiv:1901.08563 [hep-th]}}.

\bibitem{Bourjaily:2018yfy}
J.~L. Bourjaily, A.~J. McLeod, M.~von Hippel, and M.~Wilhelm, ``{A (Bounded)
  Bestiary of Feynman Integral Calabi-Yau Geometries},''
  \href{http://dx.doi.org/10.1103/PhysRevLett.122.031601}{{\em Phys. Rev.
  Lett.} {\bf 122} (2019) no. 3, 031601},
\href{http://arxiv.org/abs/1810.07689}{{ arXiv:1810.07689 [hep-th]}}.

\bibitem{Brown:2009ta}
F.~C.~S. Brown, ``{On the Periods of Some Feynman Integrals},''
\href{http://arxiv.org/abs/0910.0114}{{ arXiv:0910.0114 [math.AG]}}.

\bibitem{Panzer:2014caa}
E.~Panzer, ``{Algorithms for the Symbolic Integration of Hyperlogarithms with
  Applications to Feynman Integrals},''
  \href{http://dx.doi.org/10.1016/j.cpc.2014.10.019}{{\em Comput. Phys.
  Commun.} {\bf 188} (2015)  148--166}, \href{http://arxiv.org/abs/1403.3385}{{
  arXiv:1403.3385 [hep-th]}}.
{\tt HyperInt} is obtainable at
  \href{https://bitbucket.org/PanzerErik/hyperint/wiki/Home}{this URL}.

\bibitem{Bourjaily:2018aeq}
J.~L. Bourjaily, A.~J. McLeod, M.~von Hippel, and M.~Wilhelm, ``{Rationalizing
  Loop Integration},'' \href{http://dx.doi.org/10.1007/JHEP08(2018)184}{{\em
  JHEP} {\bf 08} (2018)  184},
\href{http://arxiv.org/abs/1805.10281}{{ arXiv:1805.10281 [hep-th]}}.

\bibitem{Bourjaily:2019igt}
J.~L. Bourjaily, A.~J. McLeod, C.~Vergu, M.~Volk, M.~von Hippel, and
  M.~Wilhelm, ``{Rooting Out Letters: Octagonal Symbol Alphabets and Algebraic
  Number Theory},''
\href{http://arxiv.org/abs/1910.14224}{{ arXiv:1910.14224 [hep-th]}}.

\bibitem{CaronHuot:2011ky}
S.~Caron-Huot, ``{Superconformal Symmetry and Two-Loop Amplitudes in Planar
  $\mathcal{N}\!=\!4$ Super Yang-Mills},''
  \href{http://dx.doi.org/10.1007/JHEP12(2011)066}{{\em JHEP} {\bf 1112} (2011)
   066},
\href{http://arxiv.org/abs/1105.5606}{{ arXiv:1105.5606 [hep-th]}}.

\bibitem{CaronHuot:2010ek}
S.~Caron-Huot, ``{Notes on the Scattering Amplitude / Wilson Loop Duality},''
  \href{http://dx.doi.org/10.1007/JHEP07(2011)058}{{\em JHEP} {\bf 1107} (2011)
   058},
\href{http://arxiv.org/abs/1010.1167}{{ arXiv:1010.1167 [hep-th]}}.

\bibitem{CaronHuot:2011kk}
S.~Caron-Huot and S.~He, ``{Jumpstarting the All-Loop $S$-Matrix of Planar
  $\mathcal{N}\!=\!4$ Super Yang-Mills},''
  \href{http://dx.doi.org/10.1007/JHEP07(2012)174}{{\em JHEP} {\bf 1207} (2012)
   174},
\href{http://arxiv.org/abs/1112.1060}{{ arXiv:1112.1060 [hep-th]}}.

\bibitem{Golden:2014xqf}
J.~Golden and M.~Spradlin, ``{An Analytic Result for the Two-Loop Seven-Point
  MHV Amplitude in $\mathcal{N}\!=\!4$ SYM},''
  \href{http://dx.doi.org/10.1007/JHEP08(2014)154}{{\em JHEP} {\bf 1408} (2014)
   154},
\href{http://arxiv.org/abs/1406.2055}{{ arXiv:1406.2055 [hep-th]}}.

\bibitem{Bourjaily:2019jrk}
J.~L. Bourjaily, F.~Dulat, and E.~Panzer, ``{Manifestly Dual-Conformal Loop
  Integration},'' \href{http://dx.doi.org/10.1016/j.nuclphysb.2019.03.022}{{\em
  Nucl. Phys.} {\bf B942} (2019)  251--302},
\href{http://arxiv.org/abs/1901.02887}{{ arXiv:1901.02887 [hep-th]}}.

\bibitem{Vergu:2009zm}
C.~Vergu, ``{Higher point MHV Amplitudes in $\mathcal{N}\!=\!4$ Supersymmetric
  Yang-Mills Theory},''
  \href{http://dx.doi.org/10.1103/PhysRevD.79.125005}{{\em Phys. Rev.} {\bf
  D79} (2009)  125005},
\href{http://arxiv.org/abs/0903.3526}{{ arXiv:0903.3526 [hep-th]}}.

\bibitem{Vergu:2009tu}
C.~Vergu, ``{The Two-Loop MHV Amplitudes in $\mathcal{N}\!=\!4$ Supersymmetric
  Yang- Mills Theory},''
  \href{http://dx.doi.org/10.1103/PhysRevD.80.125025}{{\em Phys. Rev.} {\bf
  D80} (2009)  125025},
\href{http://arxiv.org/abs/0908.2394}{{ arXiv:0908.2394 [hep-th]}}.

\bibitem{Hodges:2009hk}
A.~Hodges, ``{Eliminating Spurious Poles from Gauge-Theoretic Amplitudes},''
  \href{http://dx.doi.org/10.1007/JHEP05(2013)135}{{\em JHEP} {\bf 1305} (2013)
   135},
\href{http://arxiv.org/abs/0905.1473}{{ arXiv:0905.1473 [hep-th]}}.

\bibitem{Cheng:1987ga}
H.~Cheng and T.~T. Wu, {\em {Expanding Protons: Scattering at High Energies}}.
\newblock Cambridge, USA: MIT-PR.,
1987.
\newblock

\bibitem{Bourjaily:2012gy}
J.~L. Bourjaily, ``{Positroids, Plabic Graphs, and Scattering Amplitudes in
  {\sc Mathematica}},''
\href{http://arxiv.org/abs/1212.6974}{{ arXiv:1212.6974 [hep-th]}}.

\bibitem{Anastasiou:2013srw}
C.~Anastasiou, C.~Duhr, F.~Dulat, and B.~Mistlberger, ``{Soft Triple-Real
  Radiation for Higgs Production at N$_3$LO},''
  \href{http://dx.doi.org/10.1007/JHEP07(2013)003}{{\em JHEP} {\bf 07} (2013)
  003},
\href{http://arxiv.org/abs/1302.4379}{{ arXiv:1302.4379 [hep-ph]}}.

\bibitem{Panzer:2014gra}
E.~Panzer, ``{On Hyperlogarithms and Feynman Integrals with Divergences and
  Many Scales},'' \href{http://dx.doi.org/10.1007/JHEP03(2014)071}{{\em JHEP}
  {\bf 03} (2014)  071},
\href{http://arxiv.org/abs/1401.4361}{{ arXiv:1401.4361 [hep-th]}}.

\bibitem{Brown:2008um}
F.~Brown, ``{The Massless Higher-Loop Two-Point Function},''
  \href{http://dx.doi.org/10.1007/s00220-009-0740-5}{{\em Commun. Math. Phys.}
  {\bf 287} (2009)  925--958},
\href{http://arxiv.org/abs/0804.1660}{{ arXiv:0804.1660 [math.AG]}}.

\bibitem{Besier:2018jen}
M.~Besier, D.~van Straten, and S.~Weinzierl, ``{Rationalizing Roots: an
  Algorithmic Approach},''
\href{http://arxiv.org/abs/1809.10983}{{ arXiv:1809.10983 [hep-th]}}.

\bibitem{Arkani-Hamed:2014via}
N.~Arkani-Hamed, J.~L. Bourjaily, F.~Cachazo, and J.~Trnka, ``{Singularity
  Structure of Maximally Supersymmetric Scattering Amplitudes},''
  \href{http://dx.doi.org/10.1103/PhysRevLett.113.261603}{{\em Phys. Rev.
  Lett.} {\bf 113} (2014) no. 26, 261603},
\href{http://arxiv.org/abs/1410.0354}{{ arXiv:1410.0354 [hep-th]}}.

\bibitem{Duhr:2011zq}
C.~Duhr, H.~Gangl, and J.~R. Rhodes, ``{From Polygons and Symbols to
  Polylogarithmic Functions},''
  \href{http://dx.doi.org/10.1007/JHEP10(2012)075}{{\em JHEP} {\bf 10} (2012)
  075},
\href{http://arxiv.org/abs/1110.0458}{{ arXiv:1110.0458 [math-ph]}}.

\bibitem{Alday:2009zm}
L.~F. Alday, J.~M. Henn, J.~Plefka, and T.~Schuster, ``{Scattering into the
  Fifth Dimension of $\mathcal{N}\!=\!4$ super Yang-Mills},''
  \href{http://dx.doi.org/10.1007/JHEP01(2010)077}{{\em JHEP} {\bf 1001} (2010)
   077},
\href{http://arxiv.org/abs/0908.0684}{{ arXiv:0908.0684 [hep-th]}}.

\bibitem{Henn:2010ir}
J.~M. Henn, S.~G. Naculich, H.~J. Schnitzer, and M.~Spradlin, ``{More Loops and
  Legs in Higgs-Regulated $\mathcal{N}\!=\!4$ SYM Amplitudes},''
  \href{http://dx.doi.org/10.1007/JHEP08(2010)002}{{\em JHEP} {\bf 08} (2010)
  002},
\href{http://arxiv.org/abs/1004.5381}{{ arXiv:1004.5381 [hep-th]}}.

\bibitem{Bern:2005iz}
Z.~Bern, L.~J. Dixon, and V.~A. Smirnov, ``{Iteration of Planar Amplitudes in
  Maximally Supersymmetric Yang-Mills Theory at Three Loops and Beyond},''
  \href{http://dx.doi.org/10.1103/PhysRevD.72.085001}{{\em Phys. Rev.} {\bf
  D72} (2005)  085001},
\href{http://arxiv.org/abs/hep-th/0505205}{{ arXiv:hep-th/0505205}}.

\bibitem{Sterman:2002qn}
G.~F. Sterman and M.~E. Tejeda-Yeomans, ``{Multiloop Amplitudes and
  Resummation},'' \href{http://dx.doi.org/10.1016/S0370-2693(02)03100-3}{{\em
  Phys. Lett.} {\bf B552} (2003)  48--56},
\href{http://arxiv.org/abs/hep-ph/0210130}{{ arXiv:hep-ph/0210130 [hep-ph]}}.

\bibitem{Beisert:2006ez}
N.~Beisert, B.~Eden, and M.~Staudacher, ``{Transcendentality and Crossing},''
  \href{http://dx.doi.org/10.1088/1742-5468/2007/01/P01021}{{\em J. Stat.
  Mech.} {\bf 0701} (2007)  P01021},
\href{http://arxiv.org/abs/hep-th/0610251}{{ arXiv:hep-th/0610251 [hep-th]}}.

\bibitem{Correa:2012nk}
D.~Correa, J.~Henn, J.~Maldacena, and A.~Sever, ``{The Cusp Anomalous Dimension
  at Three Loops and Beyond},''
  \href{http://dx.doi.org/10.1007/JHEP05(2012)098}{{\em JHEP} {\bf 05} (2012)
  098},
\href{http://arxiv.org/abs/1203.1019}{{ arXiv:1203.1019 [hep-th]}}.

\bibitem{Henn:2013wfa}
J.~M. Henn and T.~Huber, ``{The Four-Loop Cusp Anomalous Dimension in
  $\mathcal{N}\!=\!4$ Super Yang-Mills and Analytic Integration Techniques for
  Wilson Line Integrals},''
  \href{http://dx.doi.org/10.1007/JHEP09(2013)147}{{\em JHEP} {\bf 09} (2013)
  147},
\href{http://arxiv.org/abs/1304.6418}{{ arXiv:1304.6418 [hep-th]}}.

\bibitem{Kotikov:2004er}
A.~V. Kotikov, L.~N. Lipatov, A.~I. Onishchenko, and V.~N. Velizhanin, ``{Three
  Loop Universal Anomalous Dimension of the Wilson Operators in
  $\mathcal{N}\!=\!4$ SUSY Yang-Mills Model},''
  \href{http://dx.doi.org/10.1016/j.physletb.2004.05.078,
  10.1016/j.physletb.2005.11.002}{{\em Phys. Lett.} {\bf B595} (2004)
  521--529}, \href{http://arxiv.org/abs/hep-th/0404092}{{ arXiv:hep-th/0404092
  [hep-th]}}.
[Erratum: Phys. Lett.B632,754(2006)].

\bibitem{Freyhult:2007pz}
L.~Freyhult, A.~Rej, and M.~Staudacher, ``{A Generalized Scaling Function for
  AdS/CFT},'' \href{http://dx.doi.org/10.1088/1742-5468/2008/07/P07015}{{\em J.
  Stat. Mech.} {\bf 0807} (2008)  P07015},
\href{http://arxiv.org/abs/0712.2743}{{ arXiv:0712.2743 [hep-th]}}.

\bibitem{Golden:2013lha}
J.~Golden and M.~Spradlin, ``{The Differential of All Two-Loop MHV Amplitudes
  in $\mathcal{N}\!=\!4$ Yang-Mills Theory},''
  \href{http://dx.doi.org/10.1007/JHEP09(2013)111}{{\em JHEP} {\bf 1309} (2013)
   111},
\href{http://arxiv.org/abs/1306.1833}{{ arXiv:1306.1833 [hep-th]}}.

\bibitem{Golden:2013xva}
J.~Golden, A.~B. Goncharov, M.~Spradlin, C.~Vergu, and A.~Volovich, ``{Motivic
  Amplitudes and Cluster Coordinates},''
  \href{http://dx.doi.org/10.1007/JHEP01(2014)091}{{\em JHEP} {\bf 1401} (2014)
   091},
\href{http://arxiv.org/abs/1305.1617}{{ arXiv:1305.1617 [hep-th]}}.

\bibitem{Golden:2014pua}
J.~Golden and M.~Spradlin, ``{A Cluster Bootstrap for Two-Loop MHV
  Amplitudes},'' \href{http://dx.doi.org/10.1007/JHEP02(2015)002}{{\em JHEP}
  {\bf 1502} (2015)  002},
\href{http://arxiv.org/abs/1411.3289}{{ arXiv:1411.3289 [hep-th]}}.

\bibitem{Golden:2014xqa}
J.~Golden, M.~F. Paulos, M.~Spradlin, and A.~Volovich, ``{Cluster
  Polylogarithms for Scattering Amplitudes},''
  \href{http://dx.doi.org/10.1088/1751-8113/47/47/474005}{{\em J. Phys.} {\bf
  A47} (2014) no. 47, 474005},
\href{http://arxiv.org/abs/1401.6446}{{ arXiv:1401.6446 [hep-th]}}.

\bibitem{Drummond:2017ssj}
J.~Drummond, J.~Foster, and {\"O}.~G{\"u}rdo\v{g}an, ``{Cluster Adjacency
  Properties of Scattering Amplitudes},''
\href{http://arxiv.org/abs/1710.10953}{{ arXiv:1710.10953 [hep-th]}}.

\end{thebibliography}
\end{document}